\documentclass[traditabstract]{aa} 

\usepackage{graphicx}
\usepackage{txfonts}
\usepackage{natbib}
\usepackage{ gensymb }
\usepackage{setspace}
\usepackage{subfigure}
\usepackage{dcolumn}
\usepackage{booktabs}
\usepackage{xfrac}
\usepackage{color}

\begin{document}

   \titlerunning{A low-mass, pre-main sequence eclipsing binary with evidence of a circumbinary disk}

   \title{CoRoT\thanks{The CoRoT space mission was developed and is operated by the French space agency CNES, with participation of ESAÕs RSSD and Science Programmes, Austria, Belgium, Brazil, Germany, and Spain.}\,223992193: A new, low-mass, pre-main sequence eclipsing binary with evidence of a circumbinary disk}

   \author{E. Gillen \inst{1} \fnmsep \thanks{email: ed.gillen@astro.ox.ac.uk}
          \and S. Aigrain \inst{1} 
          \and A. McQuillan \inst{1,2} 
          \and J. Bouvier \inst{3} 
          \and S. Hodgkin \inst{4} 
          \and S. H. P. Alencar \inst{5} 
          \and C. Terquem \inst{1}
          \and J. Southworth \inst{6} 
          \and N. P. Gibson \inst{7}
           \and A. Cody \inst{8}
          \and M. Lendl \inst{9} 
          \and M. Morales-Calder\'{o}n \inst{10}
          \and F. Favata \inst{11}
          \and J. Stauffer \inst{8} 
          \and G. Micela \inst{12} 
          }

   \institute{Sub-department of Astrophysics, Department of Physics, University of Oxford, Keble Road, Oxford OX1 3RH, UK 
         \and School of Physics and Astronomy, Raymond and Beverly Sackler, Faculty of Exact Sciences, Tel Aviv University, 69978, Tel Aviv, Israel
	\and UJF-Grenoble 1 / CNRS-INSU, Institut de Plan\'etologie et d'Astrophysique de Grenoble (IPAG) UMR 5274, Grenoble, F-38041, France
	\and Institute of Astronomy, Madingley Road, Cambridge CB3 0HA
         \and Departamento de F\'{i}sica - ICEx - UFMG, Av. Ant\^{o}nio Carlos, 6627, 30270-901, Belo Horizonte, MG, Brazil
         \and Astrophysics Group, Keele University, Staffordshire, ST5 5BG, UK
         \and European Southern Observatory, Karl-Schwarzschild-Str. 2, 85748 Garching bei M\"{u}nchen, Germany
         \and Spitzer Science Center, California Institute of Technology, 1200 E California Blvd., Pasadena, CA 91125, USA
         \and Observatoire de Gen\`{e}ve, Universit\'{e} de Gen\`{e}ve, Chemin des maillettes 51, 1290 Sauverny, Switzerland
         \and Centro de Astrobiolog\'{i}a (INTA-CSIC); ESAC Campus, P.O. Box 78, E-28691 Villanueva de la Canada, Spain
         \and European Space Agency, 8-10 rue Mario Nikis, 75738 Paris Cedex 15, France
         \and INAF Ð Osservatorio Astronomico di Palermo, Piazza del Parlamento 1, 90134, Palermo, Italy}

   \date{Received \ldots; accepted \ldots}

   \abstract{We present the discovery of CoRoT\,223992193, a double-lined, detached eclipsing binary, comprising two pre-main sequence M dwarfs, discovered by the CoRoT space mission during a 23-day observation of the 3 Myr old NGC\,2264 star-forming region. Using multi-epoch optical and near-IR follow-up spectroscopy with FLAMES on the Very Large Telescope and ISIS on the William Herschel Telescope we obtain a full orbital solution and derive the fundamental parameters of both stars by modelling the light curve and radial velocity data. The orbit is circular and has a period of $3.8745745 \pm 0.0000014$ days. The masses and radii of the two stars are $0.67 \pm 0.01$ and $0.495 \pm 0.007$ $M_{\sun}$ and $1.30 \pm 0.04$ and $1.11 ~^{+0.04}_{-0.05}$ $R_{\sun}$, respectively. This system is a useful test of evolutionary models of young low-mass stars, as it lies in a region of parameter space where observational constraints are scarce; comparison with these models indicates an apparent age of $\sim$3.5--6\,Myr. The systemic velocity is within $1\sigma$ of the cluster value which, along with the presence of lithium absorption, strongly indicates cluster membership. The CoRoT light curve also contains large-amplitude, rapidly evolving out-of-eclipse variations, which are difficult to explain using starspots alone. 
The system's spectral energy distribution reveals a mid-infrared excess, which we model as thermal emission from a small amount of dust located in the inner cavity of a circumbinary disk. In turn, this opens up the possibility that some of the out-of-eclipse variability could be due to occultations of the central stars by material located at the inner edge or in the central cavity of the circumbinary disk.}

  
   \maketitle


\section{Introduction}

\begin{figure*}
  \centering  
  \includegraphics[width=0.8\linewidth]{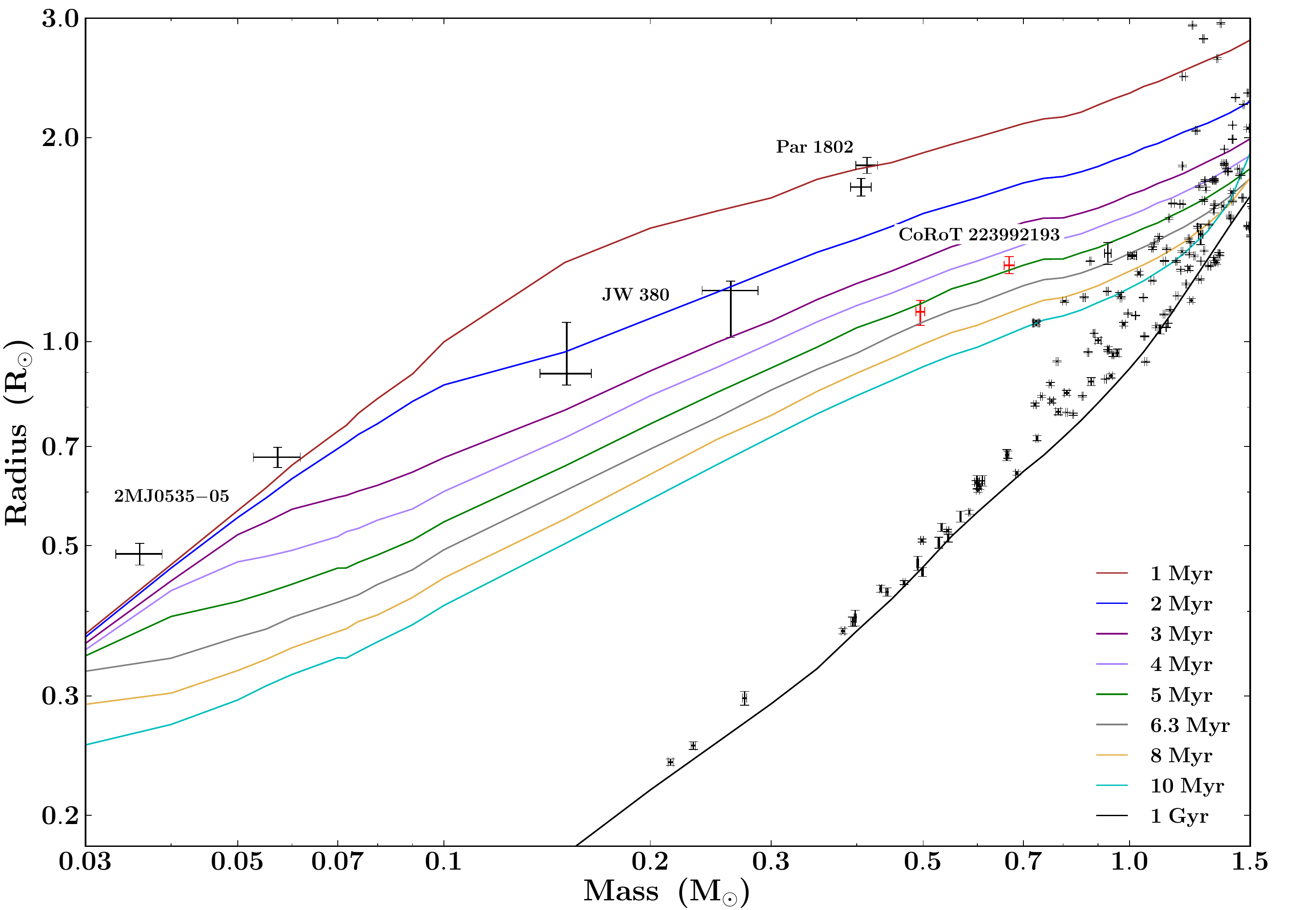}
   \caption{Mass-radius relation for low-mass EBs. The black points show measurements for stars with masses $< 1.5\,M_{\sun}$ in detached EBs$^{1}$, and the lines show, from top to bottom, the theoretical isochrones of \protect\citet[][BCAH98]{Baraffe98} for 1, 2, 3, 4, 5, 6.3, 8, 10\,Myr and 1\,Gyr (brown, blue, purple, lilac, green, grey, ochre, cyan and black, respectively. $Y=0.282$, $[{\rm M}/{\rm H}]=0$, mixing length $\alpha = 1.9$). The components of the new system presented in this paper are shown in red. Note that it lies in a very sparsely populated region of the diagram, making it a valuable test of PMS stellar evolution models. For comparison, we have also labelled the three lowest mass systems known in the Orion Nebula Cluster (see Table~\protect\ref{pms_ebs} for details. For clarity, the higher mass systems are not labelled).}
   \label{MR_plot}
\end{figure*}
 
Detached, double-lined eclipsing binaries (EBs) are extremely valuable objects because their masses, radii, effective temperatures and luminosities can be determined in a model-independent manner from the light and radial velocity curves of the system. When these reach a precision of a few percent or less, they provide one of the most powerful tests of stellar evolution models available \citep{Andersen91,Torres10}. As these models underpin much of astrophysics, it is vital that they are tested as rigorously as possible. The two components of a given EB can generally be assumed to share the same age and metallicity, which adds to the tightness of the constraints. 

Figure \ref{MR_plot} shows the existing mass and radius measurements for low-mass, detached EBs\footnote{\label{ftebs}Data from John Southworth's catalogue, \tt{http://www.astro.keele.ac.uk/$\sim$jkt/debdata/debs.html}.}.  
While there are now many well-characterised systems on the main sequence, there are very few on the pre-main sequence (PMS): to the best of our knowledge, there are only eight published low-mass EBs (with known stellar parameters), where both components are PMS objects with masses below $1.5\,M_{\odot}$ (see Table~\ref{pms_ebs}), and all but one of these are located in Orion. The PMS is an important region of parameter space because it corresponds to a period of very rapid evolution, when the models are still sensitive to their initial conditions, motivating efforts to discover and characterise more of these systems. 

\begin{table*}
  \centering
  \caption[]{Known low-mass, pre-main sequence eclipsing binary systems.}
  \label{pms_ebs}
  \begin{tabular}{llllllll}
    \hline
    \noalign{\smallskip}
    Name & $M_{\rm{pri}}$ & $M_{\rm{sec}}$ & $R_{\rm{pri}}$ & $R_{\rm{sec}}$ & Cluster\,* & Age & Reference(s) \\
    & ($M_\odot$) & ($M_\odot$) & ($R_\odot$) & ($R_\odot$) & & (Myr) & \\
    \noalign{\smallskip}
    \hline
    \noalign{\smallskip}
    2MJ0535-05 & $0.06$ & $0.04$ & $0.68$ &  $0.49$ & ONC & $\sim 1$ & \citet{Stassun06,Stassun07} \\
    JW\,380 & $0.26$ & $ 0.15$ & $1.19$ &  $0.90$ & ONC & $\sim 1$ & \citet{Irwin07} \\
    Par\,1802 & $0.41$ & $0.41$ & $1.82$ &  $1.69$ & ONC & $\sim 1$ & \citet{Cargile08,Stassun08} \\
    ISOY\,J0535-0447 & $0.83^{\rm a}$ & $0.05^{\rm a}$ &  &   & ONC & $\sim 1$ & \citet{Morales-calderon12} \\
    V1174\,Ori & $1.00$ & $0.73$ & $1.34$ &  $1.07$ & Ori OB 1c & $\sim5$--$10$ & \citet{Stassun04} \\ 
    RXJ\,0529.4$+$0041A & $1.27$ & $0.93$ & $1.44$ &  $1.35$ & Ori OB 1a & $\sim 7$--$13$ & \citet{Covino00,Covino01,Covino04} \\
    ASAS\,J0528$+$03 & $1.38$ & $1.33$ & $1.83$ &  $1.73$ & Ori OB 1a & $\sim 7$--$13$ & \citet{Stempels08} \\
    MML\,53 & $\geq 0.97$ & $\geq 0.84$ &  &  & UCL & $\sim15\,$ & \citet{Hebb10, Hebb11} \\
    \noalign{\smallskip}
    \hline
  \end{tabular}
  \begin{list}{}{}
  \item[$^{\mathrm{a}}$]preliminary estimates.
  \item[*]ONC = Orion Nebula Cluster and UCL = Upper Centaurus Lupus.
  \end{list}
\end{table*}

Perhaps the most important discrepancy between current observations and models is the fact that low-mass stars are observed to be cooler and larger than predicted by the models: theoretical models under-predict radii by up to 20\% \citep{Covino04,Stassun07,Coughlin11} and masses by 10 to 30\% (the greater discrepancy corresponding to lower masses; \citealt{Hillenbrand04,Mathieu07}), for both PMS and main sequence objects. Testing stellar evolution models with EBs assumes that the stars can be considered to evolve independently. This has been questioned for close-separation binaries, which are spun up by tidal interactions, and are more magnetically active than single stars of comparable masses and ages: \citet{Chabrier07} studied the impact of this enhanced magnetic field on convective processes within low-mass stars, finding that it could explain the observed cool temperatures and large radii. This is especially relevant to young binaries, which are particularly active. \citet{Coughlin11} find that the radius excess seen in members of EBs decreases as the orbital period increases, suggesting that the radius excess may be a by-product of interactions between the two components of a close binary system. Indeed, long period main sequence EBs, with component masses $M \gtrsim 0.7\,M_{\odot}$, appear not to exhibit a radius excess \citep[e.g.][]{Lacy05,Bass12,Southworth13}. However, it is worth noting that at lower masses a radius excess is observed in some main sequence systems, even at long periods \citep[e.g.][]{Irwin11}.

Focussing on binary stars for model evaluation makes it difficult to distinguish between the effects of binarity and other physics which may be missing in the models. An alternative approach, which does not suffer from this problem, is to combine interferometric angular diameter measurements and distance estimates from parallaxes to obtain radii, luminosities and effective temperatures for single stars. However, this is feasible only for very nearby, bright stars, and their mass remains unknown (unless the target is a member of a visual binary with a well characterised orbit). Asteroseismology is also an exquisite probe of stellar interiors, but while it can provide model-independent density estimates, it only gives model-dependent masses and radii. Therefore, detached, double-lined EBs remain an important observational test of stellar evolutionary models, and significant resources are dedicated to discovering and characterising them across as wide a range of mass, age and metallicity as possible. 

The lack of PMS EBs is widely recognised, and numerous observational programs have been set up to detect and characterise more systems; see for example the Monitor and YSOVAR projects \citep{Aigrain07,Morales-Calderon11}. With their wide field of view, excellent photometric precision and continuous monitoring capability over weeks or even years, the CoRoT \citep{Baglin03} and Kepler \citep{Borucki10} space missions are extremely efficient at detecting EBs (see e.g. \citealt{Prsa10}), but they normally target field main sequence stars. Fortunately, the young open cluster NGC\,2264 falls within the visibility zone of CoRoT, which enabled it to be observed continuously for over three weeks in March 2008. The resulting dataset offers an unprecedented insight into the variability of PMS stars, and was used to study accretion \citep{Alencar10}, rotation \citep{Favata10,Affer13} and PMS pulsations \citep{Zwintz11}, as well as to search for eclipsing binaries.

NGC\,2264 is a very well-studied young star forming region due to its relative proximity, well-defined membership list and low foreground extinction \citep{Dahm08}. It is the dominant component of the Mon OB1 association in the Monoceros constellation, situated in the local spiral arm at a distance of $\sim$700--900\,pc \citep[e.g.][]{Sung97,Baxter09,Sung10}. The cluster age has been estimated to be $\sim$3\,Myr old from isochrone fitting to the low-mass stellar population \citep[e.g.][]{Sung97,Park00,Rebull02} but there is an apparent dispersion of $\sim$5\,Myr from the broadened sequence of suspected members, and star formation is ongoing \citep[e.g.][]{Reipurth04,Young06,Dahm08}. Literature estimates for the total stellar population are in the region of $\sim$1000 members \citep{Dahm08}. Many new candidate members have been identified from the CoRoT observation itself and the deep CFHT $ugr$ imaging (Venuti et al. in preparation) bringing the total number of confident members to at least 1500 (A. Cody and E. Flaccomio, priv. comm). The cluster has a recessional velocity of $V=22\pm3.5\,$km\,s$^{-1}$ \citep{Furesz06}, and preliminary metallicity estimates are $[\rm{Fe}/\rm{H}] \approx -0.15$ and near-solar for other elements \citep{King00}. The mean reddening along the line of sight to the cluster is $E(B-V) = 0.06 - 0.15$\,mag \citep{Walker56,Park00,Rebull02,Sung04,Mayne08}.

The 2008 CoRoT observation of NGC\,2264 revealed several tens of new EBs, over a dozen of which may be cluster members. These will be described in more detail in future publications; here we present the discovery and characterisation of one particular system, CoRoT\,223992193. This system is also referred to as 2MASS J06414422+0925024, W6712 \citep{Sung08} and has a CSI\,2264 (Coordinated Synoptic Investigation of NGC\,2264) identifier of Mon-000256. Its light curve displays deep ($\sim$30\%) eclipses with a period of 3.87\,days, as well as significant ($\sim$15\% peak-to-peak) irregular out-of-eclipse (OOE) variability. Its optical and near-infrared colours are compatible with cluster membership, and preliminary light curve modelling and follow-up radial velocity measurements showed it to be a double-lined, near-equal mass system with a total mass of $\sim 1.2\,M_\odot$ and individual radii $>1\,R_\odot$, as expected for a PMS system. This motivated additional follow-up spectroscopy and modelling, enabling us to refine the fundamental parameters of both stars and to confirm that the systemic velocity is compatible with that of the cluster. 

The system also shows indirect evidence for a proto-planetary disk, most likely circumbinary. If confirmed, it will be the first PMS EB system found to harbour such a disk. This is particularly interesting in the light of the recent discovery of a number of circumbinary exoplanets (e.g. Kepler-16 \citealt{Doyle11}, Kepler-34 and 35 \citealt{Welsh12}, Kepler-38 \citealt{Orosz12}, Kepler-47 \citealt{Orosz12a} and Kepler-64 \citealt{Schwamb12}). In all of these systems, the orbital angular momentum of the planets is closely aligned with that of the host binary, strongly suggesting in situ formation within a circumbinary disk. Initial theoretical modelling suggests that circumbinary planets around close separation binaries should be a common occurrence \citep{Alexander12}.

CoRoT\,223992193 is a very rich system. This paper focuses on the first step in its characterisation, namely the parameters of the two stars based on the 2008 CoRoT data and follow-up radial velocities. The detailed study of the interaction of the stars with their environment requires additional, multi-band photometry and spectroscopy, and will be the subject of a later paper. In Section 2, we give details of the CoRoT observations and follow up spectroscopy. In Section 3 we model the light curve and radial velocity data, deriving fundamental parameters. We discuss the properties of the stars in Section 4, where we also present a preliminary analysis of the spectral energy distribution (SED) of the system, before concluding in Section 5. 


\section{Observations}


\subsection{Photometry}
\label{photo}

CoRoT observed NCG\,2264 continuously for 23.4 days between 7 and 31 March 2008 (run SRa01). CoRoT observations are conducted in a broad $300 - 1000\,\rm{nm}$ bandpass with a standard cadence of $512\,\rm{s}$, giving 3936 photometric data points for this run. A total of 8150 stars with magnitudes $9.5 < R < 17$ were monitored in the $1.3 \times 2.6$\,degrees field-of-view of CoRoT's exoplanet channel. About 1000 of these were previously known or suspected members of NGC\,2264. These observations\footnote{The data are publicly available from the IAS CoRoT archive: \tt{http://idoc-corot.ias.u-psud.fr/}.} represent an unprecedented photometric dataset for a young cluster, both in terms of sampling and precision. Basic aperture photometry is carried out on board the satellite, and the data were further processed by the CoRoT pipeline \citep{Auvergne09} to correct them for known instrumental effects (including background and pointing jitter).

To search for eclipses, we first applied a short-baseline running median filter to exclude outlying data points and a 1\,day baseline iterative non-linear filter \citep[see][]{Aigrain04} to remove long-term variations such as stellar activity. An automated, least-squares search for trapezoidal eclipses was then performed in the light curves of all targets, followed by visual examination of all candidates with an eclipse signal-to-noise ratio above 30. Systems with only one detected eclipse were discarded, along with spurious detections induced by hot pixels (identified from the shape of the `eclipses', and the three-colour light curves, which are available for stars with $R<15$). 

This process yielded 103 eclipsing systems, which were cross-matched with 2MASS and other published data on NGC\,2264, covering: X-rays \citep{Flaccomio97,Flaccomio06,Ramirez04}; disks \citep{Rebull02}; optical and infrared variability \citep{Makidon04,Lamm04} and spectroscopic and proper motion studies \citep{Walker56}. A 2MASS $J$ vs.\ $J-K$ colour-magnitude diagram was then used in conjunction with other membership information (where available), the spatial distribution of the targets, and a magnitude limit of $R=15.5$, to select 12 possible cluster member systems for spectroscopic follow-up. 

\begin{figure}
   \centering
   \includegraphics[width=\columnwidth]{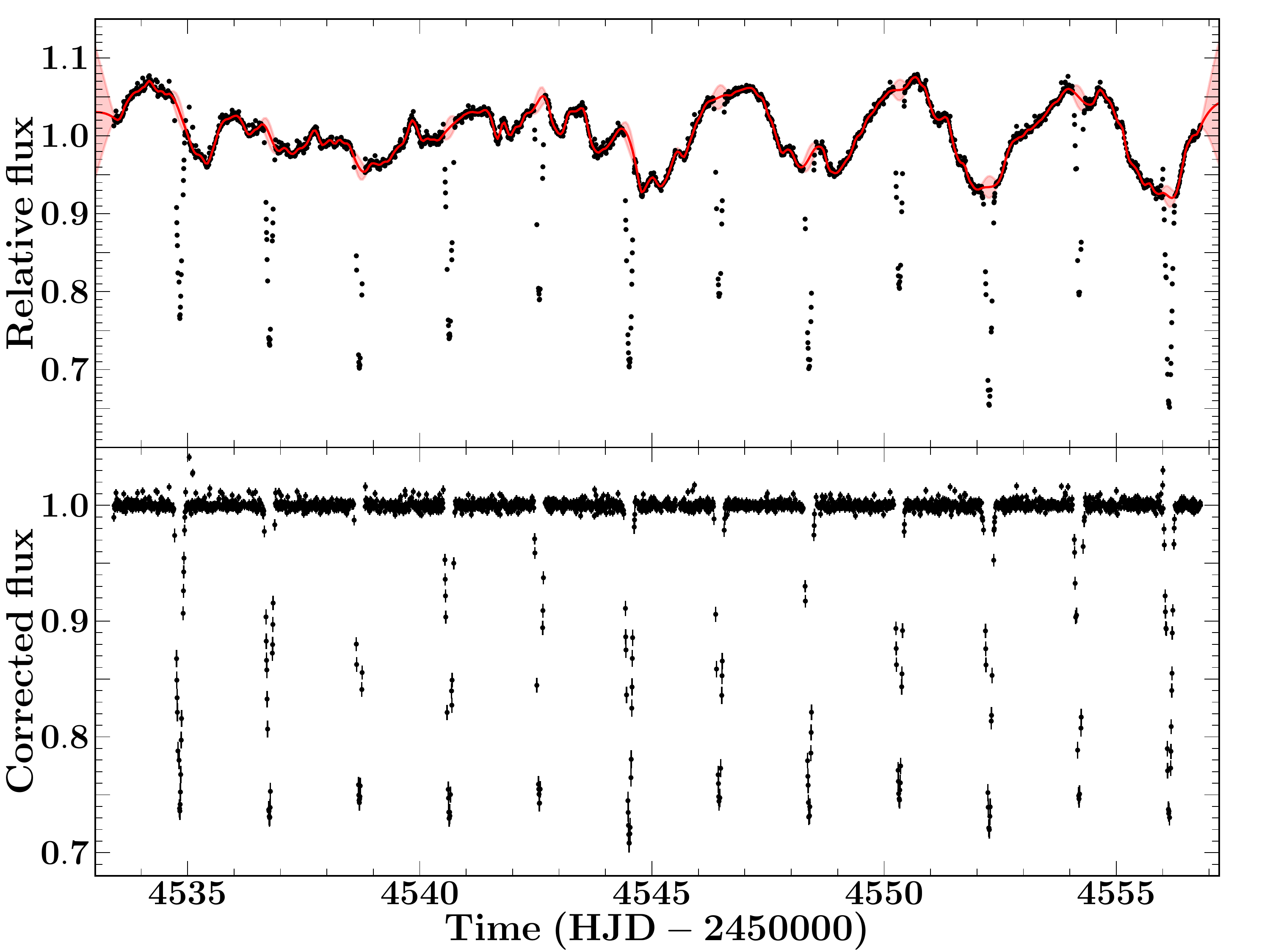}   
   \caption{Light curve of CoRoT\,223992193 obtained by CoRoT in 2008. The original light curve is shown in black in the top panel. The red line and pink shaded area show the mean and 95\% confidence interval of the predictive distribution of the Gaussian Process used to model the out-of-eclipse variations. The bottom panel shows the light curve after subtracting the mean of the predictive distribution and re-normalising to unity. The detrended light curve, which was used to model the eclipses, has a typical photometric precision of $\sim0.26\%$ per 512\,s exposure.}
   \label{ooe_rem}
\end{figure}

CoRoT\,223992193 was selected, despite being fainter than our nominal cut-off ($R=15.74$), because it fulfilled all the other criteria, and its light curve shows interesting OOE variations. The photometric properties of the system are shown in Table~\ref{eb_props}. Its magnitude and colours are consistent with a fairly low mass member of the cluster. Its light curve is shown in Figure~\ref{ooe_rem}. The large-amplitude, rapidly evolving variability seen outside the eclipses could be due to spots, but would require unusually rapid spot evolution (on timescales of hours, this is discussed in more detail in Section~\ref{spot_section}). An alternative hypothesis is that the variability might be caused by partial occultation of the stars by circumstellar or circumbinary material.

\begin{table}
  \centering
  \caption[]{Coordinates and photometric properties of CoRoT\,223992193. Note that the photometric uncertainties listed here are the formal measurement errors; they do not account for the intrinsic variability of the system.}
  \label{eb_props}
  \begin{tabular}{lll}
    \hline
    \hline
    \noalign{\smallskip}
    RA & & Dec \\
    \noalign{\smallskip}
    $06~41~44.22$ & & $+09~25~02.398$ \\
    \noalign{\smallskip}
    \hline
    \noalign{\smallskip}
    Passband & (ref.) & Magnitude \\
    \noalign{\smallskip}
    $u$ & $^{\rm a}$ & $20.021\pm0.048$ \\ 
    $g$ & $^{\rm a}$ & $17.543\pm0.005$ \\ 
    $r$ & $^{\rm a}$ & $16.090\pm0.004$ \\ 
    $i$ & $^{\rm a}$ & $15.069\pm0.004$ \\ 
    $z$ & $^{\rm a}$ & $14.479\pm0.004$\\ [1.5ex]
    $J$ & $^{\rm b}$ & $13.329\pm0.029$ \\
    $H$ & $^{\rm b}$ & $12.614\pm0.022$ \\
    $K$ & $^{\rm b}$ & $12.331\pm0.029$ \\ [1.5ex]
    $[3.6]$ & $^{\rm c}$ & $11.731\pm0.031$ \\
    $[4.5]$ & $^{\rm c}$ & $11.533\pm0.049$ \\
    $[5.8]$ & $^{\rm c}$ & $11.336\pm0.065$ \\
    $[8.0]$ & $^{\rm c}$ & $10.951\pm0.034$ \\  
    \noalign{\smallskip}
    \hline
  \end{tabular}
  \begin{list}{}{}
  \item[$^{\rm a}$]SDSS \citep{Abazajian09,Adelman-McCarthy09}, AB magnitudes
  \item[$^{\rm b}$]2MASS \citep{Cutri03}, Vega magnitudes
  \item[$^{\rm c}$]\emph{Spitzer}/IRAC \citep{Sung09}, Vega magnitudes
  \end{list}
\end{table}

\begin{figure}
   \centering
   \includegraphics[width=\columnwidth]{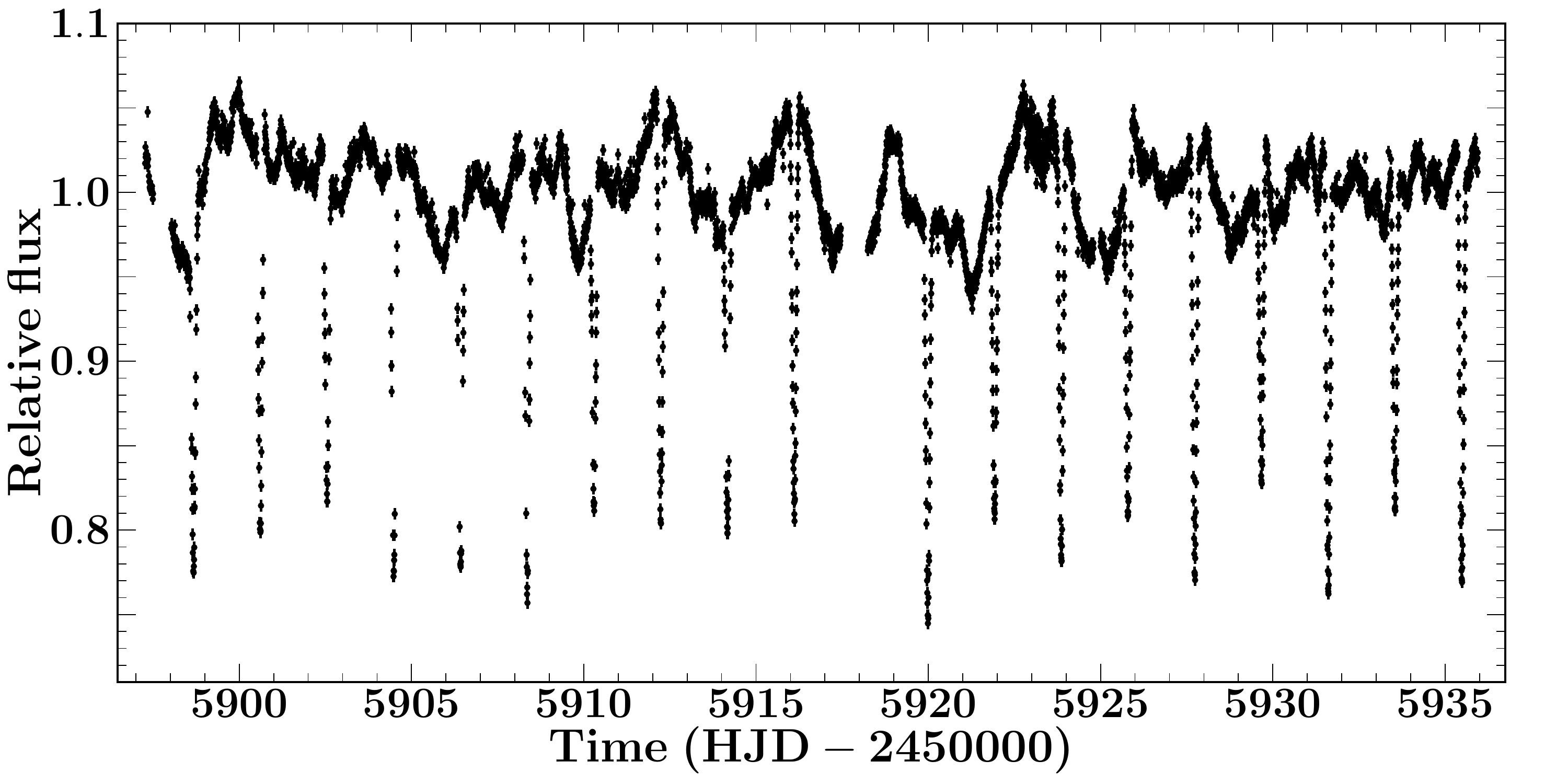}
   \caption{Light curve of CoRoT\,223992193 obtained by CoRoT in December 2011/ January 2012.}
   \label{2011lc}
\end{figure}

A second set of CoRoT observations, lasting 40 days, took place in December 2011/January 2012 (run SRa05, see Figure \ref{2011lc}). We were not able to combine the two CoRoT light curves directly, because the background appears to be significantly more spatially variable during the second run, and the current CoRoT pipeline performs only a globally averaged background correction. Improvements to the pipeline are being tested, which will hopefully resolve this problem. In the mean time, we opted to work mainly with the 2008 light curve, using the 2011/2012 data only to refine the ephemeris of the system (see Section~\ref{jktebop})


\subsection{Spectroscopy}
\label{spectro}

We performed low resolution spectroscopy to infer the combined spectral type and medium resolution spectroscopy to extract radial velocities.

\begin{figure}
\begin{center}
\includegraphics[width=\linewidth]{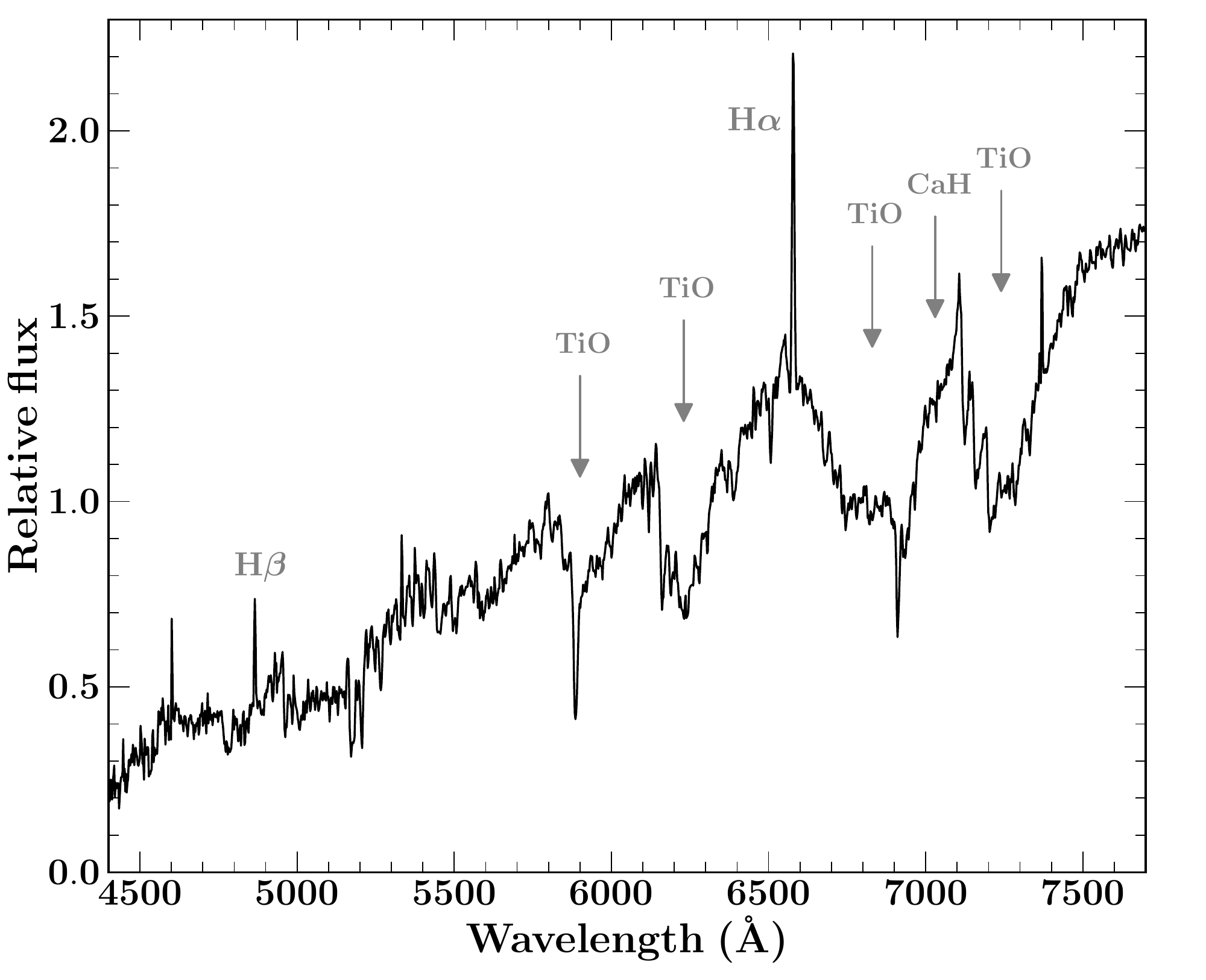}
\caption{Low-resolution optical spectrum of CoRoT\,223992193 taken with the 2.2\,m CAHA telescope. Note the strong molecular bandheads (from TiO) characteristic of the early-M combined spectral type of the target, as well as strong H$\alpha$ and H$\beta$ emission.}
\label{1039_spec}
\end{center}
\end{figure}

We obtained a low-resolution ($\sim$7\,\AA) optical spectrum of CoRoT\,223992193 on 13 April 2011 (JD 2455665.3334) using the CAFOS focal reducer \citep{Meisenheimer94} on the 2.2\,m Calar Alto telescope equipped with the G-100 grism and the SITe\#1d\_15 CCD camera, over the $4600$--$7700\,$\AA\ wavelength range. The spectrum was acquired at an airmass of 1.4 using a slit width of 1.5\,arcsec, and an exposure time of 1800\,s.  A spectrum of the spectrophotometric standard, Feige\,34, was obtained immediately after, at an airmass of 1.0, with the same set-up. The spectra of the two objects were reduced in the same way, namely: the 2D images were bias and flat-field corrected, and the stellar and neighbouring sky spectra were extracted using the {\sc iraf/twodspec} package\,\footnote{{\sc iraf} is distributed by the National Optical Astronomy Observatories, which are operated by the Association of Universities for Research in Astronomy, Inc., under cooperative agreement with the National Science Foundation.} \citep{Tody93}. The wavelength scale was calibrated using the HgHeRb spectral lamp, and owing to the small number of lines available in the arc spectrum over the observed wavelength range, the wavelength calibration is accurate to only about $20\,$\AA. Finally, the response of the instrument was corrected for in the target spectrum using that of the spectrophotometric standard. The resulting spectrum is shown in Figure~\ref{1039_spec}. It exhibits clear emission at H$\beta$ and H$\alpha$, with equivalent widths of 3.2 and 5.6\,\AA, respectively. The strong TiO absorption bands are indicative of a late spectral type. Comparing CoRoT\,223992193's photospheric spectrum with a grid of young spectral standards from \citet{AlvesdeOliveira12}, we derive a spectral type of M2 and negligible visual extinction ($A_{V} \simeq 0$). 

\begin{figure}
\begin{center}
\includegraphics[width=\linewidth]{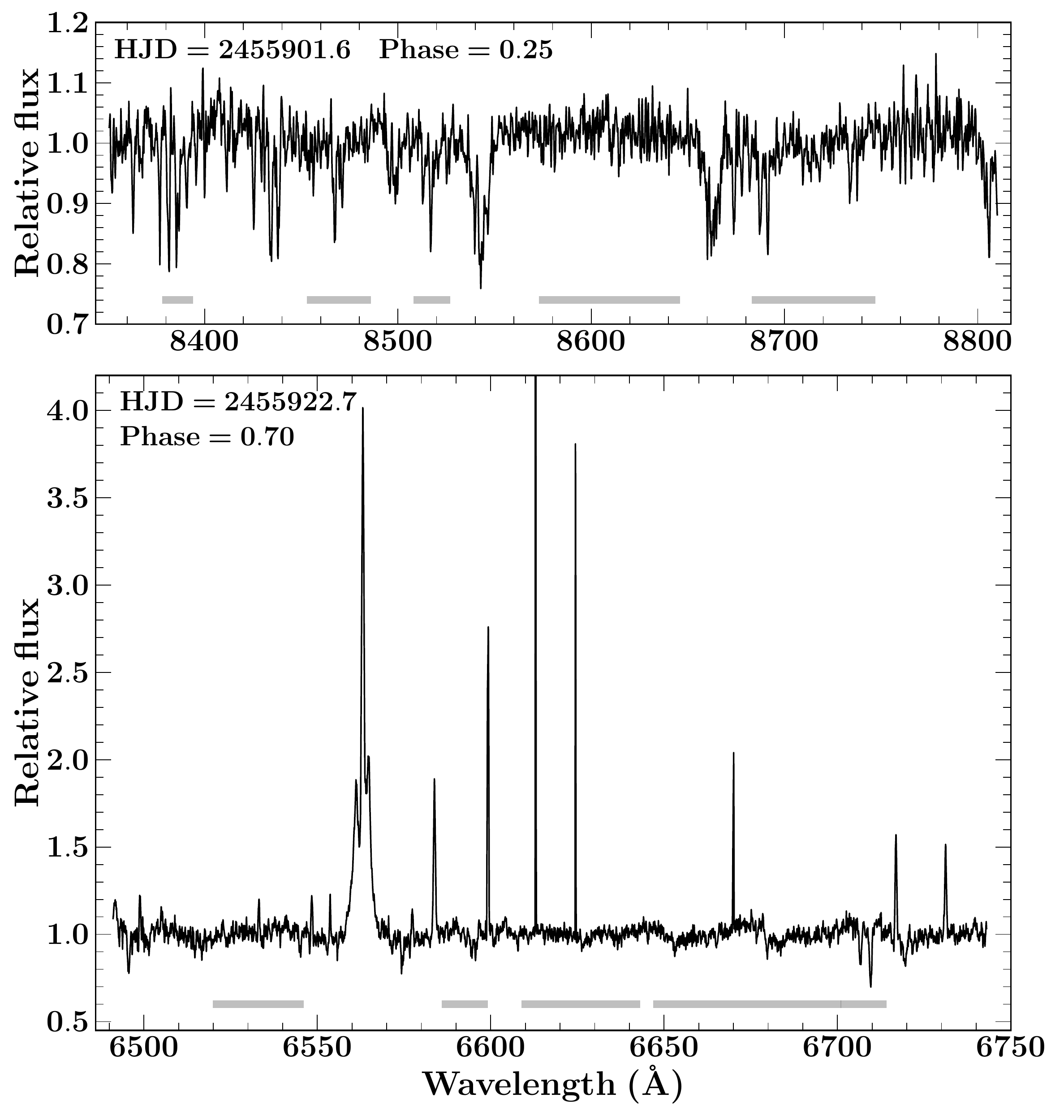}
\caption{Example continuum-corrected spectra from ISIS (top), and FLAMES (bottom). Grey shaded regions indicate the wavelength ranges used when cross-correlating the spectra with theoretical templates to measure radial velocities. Note that some of the emission lines visible in the FLAMES spectra are caused by the background nebula.}
\label{spec_each}
\end{center}
\end{figure}

We then obtained moderate resolution, near-IR spectra from the Intermediate dispersion Spectrograph and Imaging System (ISIS) on the 4\,m William Herschel Telescope (WHT) at Roque de Los Muchachos observatory on La Palma, Spain. We observed the target at seven epochs with WHT/ISIS between $3$ and $5$ December 2011, covering the wavelength range, $\sim$7850--8900\,\AA, with a spectral resolution of $R \sim$12000 and using the R1200R grating. Six observations were at quadrature and one close to eclipse. At each epoch we took three successive spectra, with exposure times of $300$ or $600\,\rm{s}$, directly followed by CuAr and CuNe arc lamp spectra. An example ISIS spectrum is shown in the top panel of Figure \ref{spec_each}. The ISIS spectra were processed with {\sc iraf}, using the \textsc{ccdproc} packages to perform the standard CCD reduction steps (trimming and overscan subtraction, bias subtraction and flat fielding) and \textsc{specred.doslit} to extract and wavelength calibrate the spectra using the arc lamp spectra. The three exposures taken at a given epoch were then combined to maximise the signal-to-noise ratio ($S/N$) and perform cosmic ray rejection. The resulting spectra typically have $S/N \sim$20 per pixel ($\sim$$25$ per spectral resolution element; see Table~\ref{spec_table}).

A further 21 medium resolution optical spectra were obtained with the FLAMES instrument on the Very Large Telescope (VLT) at Paranal, Chile, as part of a multi-object monitoring program primarily designed to study accretion (GO program 088.C-0239(A), PI Alencar). The spectra were obtained over a $\sim$3 month period (4 December 2011 - 24 February 2012), 
 with both sparsely and densely sampled time intervals. Telescope pointing issues resulted in significant loss of flux through the fibre during 6 of the 21 epochs, so that only 15 spectra could be used in the present analysis, but the phase coverage of the orbit is nonetheless fairly complete. Observations were taken using the standard setting HR15N, yielding a resolving power $R \sim17000$ over the 
wavelength range $\sim$$6440$--$6820\,$\AA. The FLAMES spectra were reduced, extracted and wavelength-calibrated using the 
 standard ESO pipeline\footnote{http://www.eso.org/sci/software/pipelines/giraffe/giraf-pipe-recipes.html}. An example FLAMES spectrum is shown in the bottom panel of Figure \ref{spec_each}. The most prominent absorption feature is the Li line at 6708.9\,\AA, which is consistent with the youth of the system. In this work, the FLAMES spectra were used only to derive radial velocities, but they contain a wealth of additional information, particularly in the broad, resolved and highly variable H$\alpha$ emission line, which will be studied in more detail in a future paper.

Finally we also obtained three spectra with the Intermediate Dispersion Spectrograph (IDS) on the 2.5\,m Isaac Newton Telescope (INT), also on La Palma. The IDS spectra cover the wavelength range $\sim$$7650$--$9300\,$\AA\ at $R \sim9300$, and were obtained between 7 and 12 March 2012. They were reduced, extracted and wavelength calibrated in the same way as the ISIS spectra. The ISIS and FLAMES spectra already gave good phase coverage of the orbit. With IDS, we obtained one spectrum at quadrature and one during each eclipse in the hope of disentangling the spectra of the two components. However, the signal-to-noise ratio of the IDS spectra turned out to be insufficient for this purpose, so we did not include them in the rest of the analysis.


\section{Analysis}


\subsection{Light Curve Modelling}


\subsubsection{Out of eclipse variability removal}
\label{ooe_removal}

The target displays significant out-of-eclipse (OOE) flux variations, which must be accounted for when modelling the CoRoT light curve to determine the fundamental parameters of the two stars. Given the complex nature of the OOE variations, it is not practical to model them at the same time as the eclipses. We therefore opted to remove the OOE variations before modelling the residuals with standard eclipse modelling software. We first tried fitting cubic splines to the OOE data (masking the in-eclipse regions) and interpolating across the eclipses. This gives fairly satisfactory results, but there is no principled way to propagate uncertainties arising from the OOE removal process through to the residuals. We therefore implemented a second approach, based on Gaussian Process (GP) regression, which provides both a better fit to the OOE data, and a more robust means of interpolating across the eclipses. 

A detailed description of GP regression is beyond the scope of the present paper; we refer the interested reader to \citet{Rasmussen06} for a textbook-level introduction, \citet{Gibson12} for a relatively detailed description in the context of astrophysical time-series data, and \citet{Aigrain12} for specific examples of GP regression applied to stellar light curves. For the present purpose, it is sufficient to think of GPs as a means of modelling the light curve by parameterising the covariance between pairs of flux measurements, rather than writing down an explicit expression for the fluxes themselves. The joint distribution of the observed fluxes is then taken to be a multi-variate Gaussian, with a covariance matrix whose elements depend on the observation times through a covariance function. After experimenting with a range of widely-used covariance functions, we opted for a member of the Mat{\'e}rn family with smoothness parameter $\nu=3/2$. This is appropriate for data displaying a relatively rough behaviour \citep{Rasmussen06}, such as we observe in the OOE light curve (see Figure \ref{ooe_rem}). The covariance $k$ between the fluxes observed at times $t$ and $t'$ is modelled as:
\begin{equation}   
k_{3/2} (r) = A^{2} \left(1+\frac{\sqrt{3}r}{l}\right) \exp\left(-\frac{\sqrt{3}r}{l}\right) + \sigma^2 \delta(r)
\end{equation}
where $r=|t-t'|$ is the time-interval between the observations and $\delta(x)$ is the Kronecker delta function. The first term represents the OOE variations, with amplitude $A$ and characteristic time scale $l$, and the second term represents white noise with standard deviation $\sigma$. For a given set of parameters, the likelihood of the model, marginalised over all the possible flux vectors which share the same covariance matrix, can be estimated analytically. It requires an inversion of the covariance matrix, which is computationally expensive, but nonetheless feasible for up to a few thousand data points.
In this analysis, we opt to subsample the data to speed up this inversion. We bin sets of $\sim$8 measurements, i.e. retaining information on much shorter timescales than the OOE variations, and take the median and mean values for the flux and time respectively. Starting with an initial guess for the OOE and white noise parameters (obtained from visual inspection of the light curve, and from photon counting statistics, respectively), we use a Nelder-Mead optimiser to find the values which maximise the likelihood. These are $A = 11.6$\,\%, $l = 1.25$ days and $\sigma = 1.5$\,mmag. We note that the white noise estimate is comparable to the estimated photon noise of $\sigma = 1.1$\,mmag for the residuals (subsampled to the same frequency as used in the OOE modelling). Once the parameters are fixed to these values, we can then compute a predictive distribution for the fluxes at any given set of times. We do this for the times of all observations (both in and out of eclipse), and obtain a light curve corrected for the OOE variations by subtracting the mean of that distribution from the observed values (Figure \ref{ooe_rem}, bottom panel). The standard deviation of the predictive distribution provides an estimate of the uncertainty on the corrected flux.

Once we have a satisfactory model for the OOE variations, we must decide whether to subtract it from the original, or divide the latter by the OOE model, before fitting the eclipses. The appropriate course of action depends on how the OOE variability affects an eclipse. For example, spots on the background star will tend to have a multiplicative effect across an eclipse, while those on the foreground star will have an additive effect. With eclipses on both stars, and with both stars likely to be spotted, the effect across all eclipses will not be the same. In addition, if some of the OOE variability arises from obscuration of one or both stars by material located outside the binary orbit (see section \ref{SED_section}), then this complicates matters further. Neither simply subtracting nor dividing will fully account for all the effects caused by the OOE variations. It may be feasible to disentangle the different effects using simultaneous light curves in widely separated bandpasses, but this is not possible using the CoRoT light curve alone. We therefore simply tested both approaches to remove the OOE variability. Subtracting gives slightly better results, in the sense that the depths of both primary and secondary eclipses are more consistent from one orbit to the next. We use the OOE-subtracted light curve when subsequently fitting the eclipses.


\subsubsection{JKTEBOP}
\label{jktebop}

Analysis of eclipsing binary light curves yields a wealth of physical information, from orbital geometry to relative stellar parameters. Light curve modelling was performed with {\sc jktebop} \citep{Southworth04,Southworth07b}, an extension of the EBOP code \citep[Eclipsing Binary Orbit Program; e.g. ][]{Popper81,Etzel81,Nelson72}. {\sc jktebop} models each star as a sphere and computes light curves through numerical integration of concentric circles over each star, employing a Levenberg-Marquardt optimisation algorithm in finding the best fit. The approximation of modelling each star as a sphere is valid in the case of well-detached systems with modest tidal distortion. 

\begin{figure}
   \centering
   \includegraphics[width=\columnwidth]{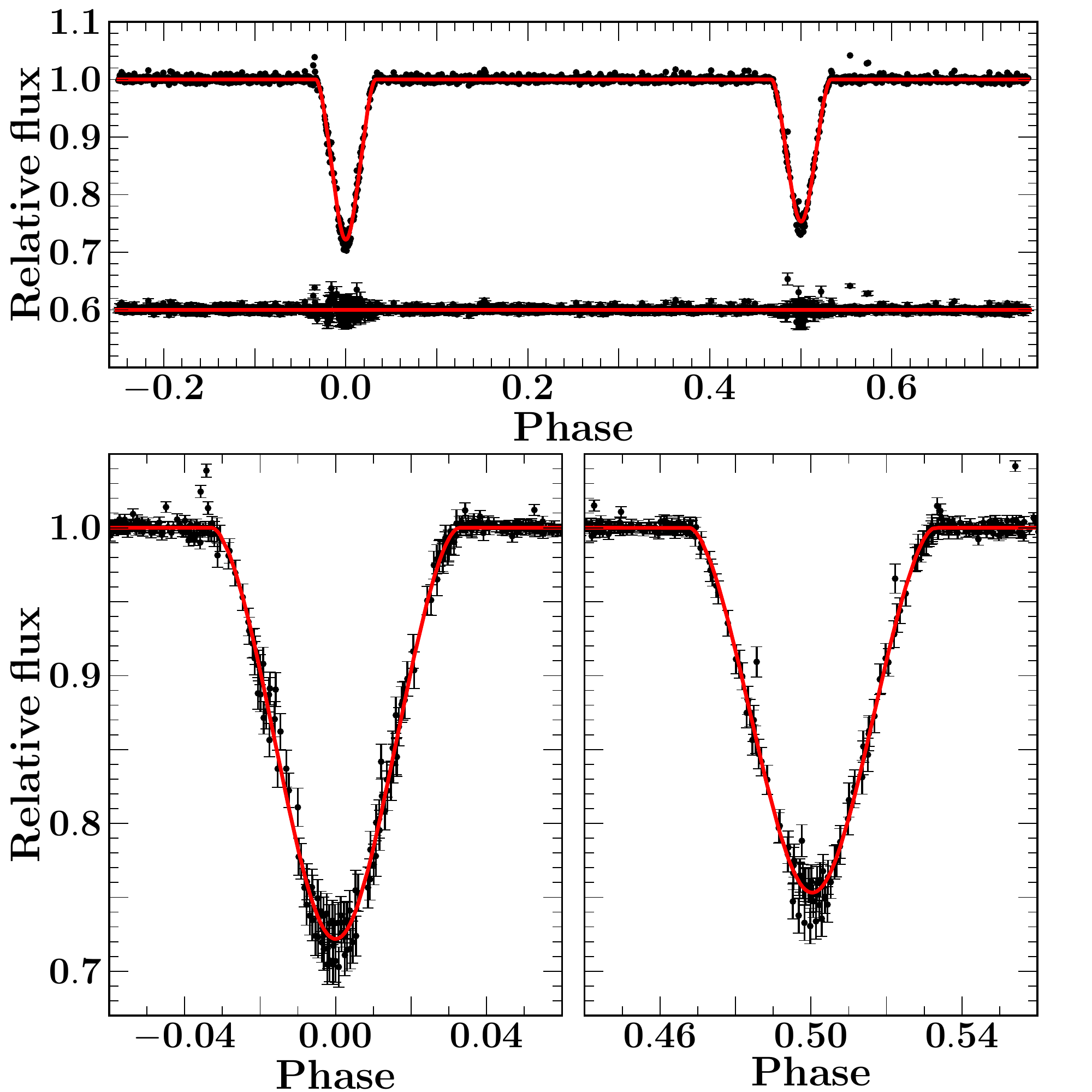}
   \caption{Top: Phase folded, detrended CoRoT light curve (black points, 2008 observation) with the {\sc jktebop} best-fit model shown in red. The residuals of the best-fit model are also shown, with a vertical offset for clarity. Phase zero marks the centre of the primary eclipse. The bottom panels show zooms on the primary and secondary eclipses (left and right, respectively).}
   \label{jktebop_fit}
\end{figure}
 
The parameters adjusted during modelling are: the central surface brightness ratio, $J = $ \textit{SB}$_{\rm{sec}}/$\textit{SB}$_{\rm{pri}}$; the sum of the radii as a fraction of the orbital separation, $( R_{\rm{pri}} + R_{\rm{sec}}) / a$; the radius ratio, $R_{\rm{sec}}/R_{\rm{pri}}$; the orbital inclination, $i$; the orbital period, $P$; the time of primary eclipse centre, $T_{\rm{prim}}$; the orbital eccentricity and the longitude of periastron, $e$ and $\omega$, in the form of the combination terms, $e \cos \omega$ and $e \sin \omega$. The initial guesses for these parameters were derived from the output of the eclipse search, from visual estimates of the ratio of eclipse depths, their durations and relative phases, and from blackbody approximations relating radius, effective temperature, luminosity and surface brightness. We checked that the final results were not sensitive to the initial guesses (provided these were sufficiently close to the final values for {\sc jktebop} to converge). Our GP treatment of the OOE variability removes reflection and gravity darkening effects from the light curve, and we therefore do not incorporate these into our modelling.

{\sc jktebop} models single-band light curves and unlike other techniques to determine eclipsing binary parameters, e.g PHOEBE \citep[Physics of Eclipsing Binaries;][]{Prsa05}, based on the Wilson-Devinney approach \citep{Wilson71}, does not require the use of model atmospheres. Effective temperatures and surface gravities therefore are only introduced in the determination of the limb darkening coefficients. We used a quadratic limb-darkening law with coefficients from ATLAS model spectra for the CoRoT bandpass \citep{Sing10}. The coefficients were specified from estimates of $T_{\rm{eff}}$, $\log g$ and $[M/H]$: effective temperatures were estimated from template spectra giving the highest stellar peaks in cross-correlation (see section \ref{RV_sec}), $\log g$ was set at $4.0$ (cgs) for both stars (reasonable for low-mass PMS stars) and metallicity was taken to be $[M/H] = -0.1$, the closest value available to the cluster metallicity \citep{King00}. We initially allowed the limb darkening parameters to vary but found that this yielded unphysical values for the parameters, while it did not significantly change the results for the other parameters. This is because the eclipses are grazing and the light curve simply does not contain enough information to constrain the limb darkening parameters. We therefore fixed them to the theoretical values in the final fit.

\begin{figure*}
   \centering
   \includegraphics[width=\linewidth]{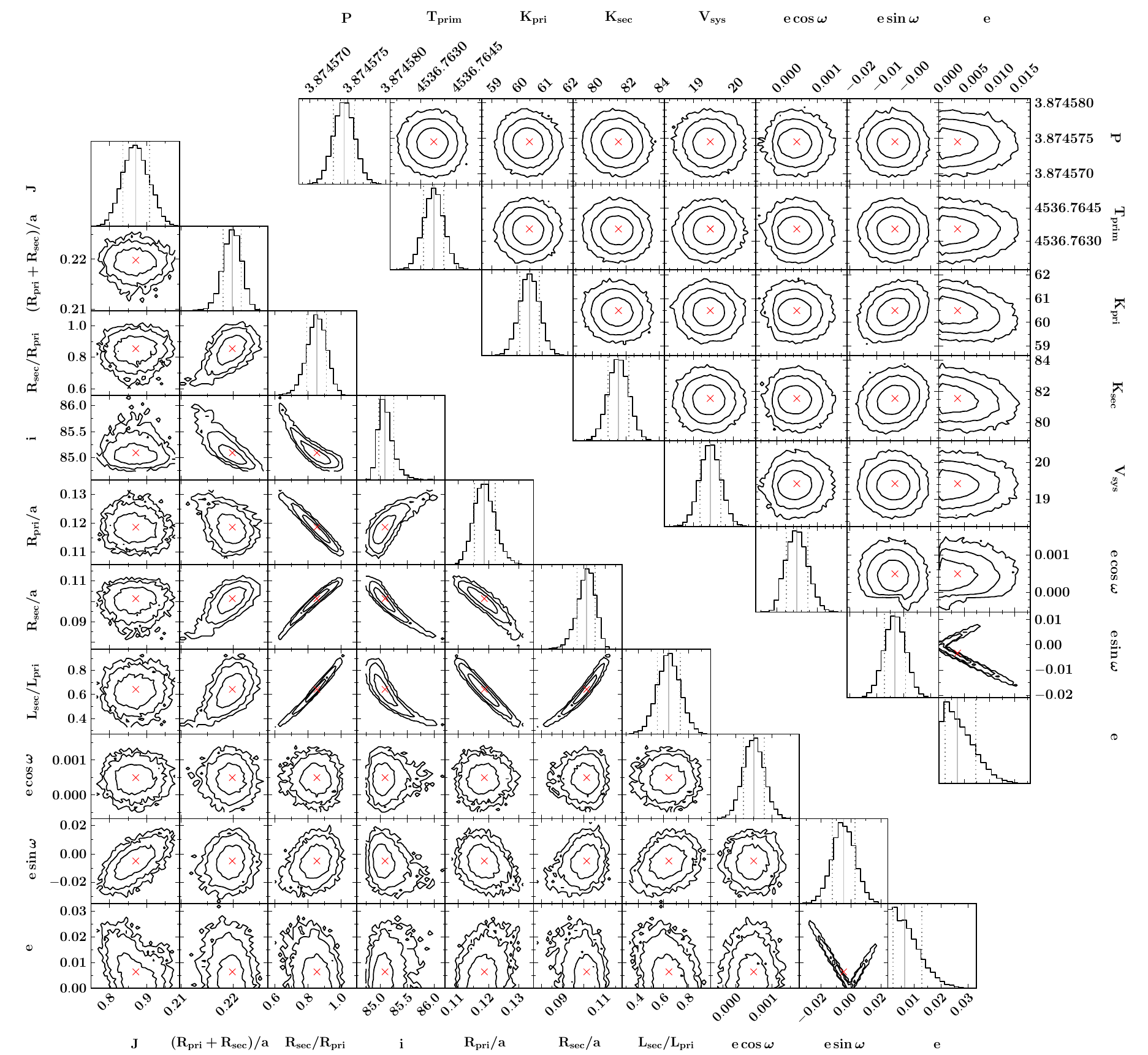}
   \caption{2-D contours and 1-D histograms of the Monte Carlo chains for selected {\sc jktebop} parameters (bottom left) and of the MCMC chains for the photometrically constrained radial velocity solution (top right; see Section \ref{RV_sec}). On the 2-D plots, red crosses show the median values and black contours represent 1, 2 and 3 $\sigma$ confidence intervals. On the histograms, solid and dashed grey vertical lines show the median and $\pm 1 \sigma$ intervals, respectively.}
    \label{conf_plot}
\end{figure*}

The fact that the eclipses are grazing also gives rise to a degeneracy between the surface brightness ratio and the radius ratio. To break this degeneracy, we added an additional constraint in the form of a light ratio, estimated from the relative cross-correlation function (CCF) peak heights. We used a FLAMES spectrum for this purpose, because the wavelength coverage of FLAMES is more similar to that of CoRoT than either ISIS or IDS, and so the relative CCF peak heights are expected to be more representative of the light ratio of the two stars in the CoRoT bandpass. We chose a single CCF (HJD = 2455940.64, see Figure \ref{gp_ccf}, left panel) with strong, well-separated, stellar peaks, and obtained a peak height ratio of 0.641 (see section \ref{RV_sec} for details of the CCF fitting procedure).

As explained in Section~\ref{photo}, we rely primarily on the 2008 light curve to model the eclipses, because the spatially dependent  background in the 2011/2012 observation affects the eclipse depths (the differences between the 2008 and the 2011/2012 depths is significantly larger than the small variation in the apparent eclipse depths during either run). However, we obtain the best-fit period and time of primary eclipse centre ($T_{\rm{prim}}$) from a combined fit to both light curves (after removing the OOE variations in the second run, in same way as for the first). The background correction error should not affect these parameters and the extended time coverage gives significantly improved accuracy.

Figure~\ref{jktebop_fit} shows the phase folded light curve along with the best-fit {\sc jktebop} model. The scatter of the residuals is significantly larger in eclipse than out of eclipse. This is partly due to occultation by the foreground star of starspots on the background star, as seen in numerous EB and transiting planet systems. In the present case, it is also due to increased uncertainty in the GP prediction across the eclipses, and to the residual variations in the depths of individual eclipses, already noted in Section~\ref{ooe_removal}, and which could be caused by changes in the global spot coverage of either star between one eclipse and the next. To account for this increased in-eclipse scatter, we performed a preliminary {\sc jktebop} run on the detrended, 2008 light curve (Figure \ref{ooe_rem}, bottom panel), and used the resulting fit to compute the reduced $\chi^{2}$ for the primary, secondary and out-of eclipse data separately. The photometric uncertainties on the fluxes were then rescaled so as to give a reduced $\chi^{2}$ of unity in each subset; the rescaling factors were 1.55 (primary), 2.68 (secondary) and 1.07 (out-of eclipse). {\sc jktebop} was then re-run (with the same initial guesses), using the scaled photometric errors.

The resulting best-fit parameters are given in Table~\ref{params}, along with uncertainties derived from a Monte Carlo analysis. This involves generating a model light curve from the best-fit parameters, adding Gaussian white noise (matching the observational errors), fitting the result, and repeating the procedure 10\,000 times to obtain distributions for each parameter, which are shown in Figure~\ref{conf_plot}. This procedure highlights the degeneracies between some parameters, most notably the radii, luminosity ratio and inclination, as expected for a grazing system, and ensures they are accounted for in the reported uncertainties. Interestingly, the use of the light ratio constraint from the FLAMES spectra enables us to break the usual degeneracy between the radius and surface brightness ratios, which is often the limiting factor in the final radius estimates for near equal-mass EBs (see e.g. the case of JW\,380, \citealt{Irwin07}). We note that the sum of the radii is roughly a fifth of the orbital separation and that the expected oblateness of each star is 0.0025 (primary) and 0.0016 (secondary), well within the allowable range for {\sc jktebop} (the EBOP model breaks down around an oblateness of 0.04, \citealt{Popper81}). This validates our initial assumption that the stars are well-detached and can be treated as spherical. The best-fit model has an eccentricity consistent with zero at $<1.5\sigma$ level.


\subsection{Radial Velocities}
\label{RV_sec}

We measured radial velocities by cross-correlating the spectra described in Section~\ref{spectro} with MARCS theoretical model spectra \citep{Gustafsson08}. 

\begin{figure*}
\centering
\includegraphics[width=0.49\linewidth]{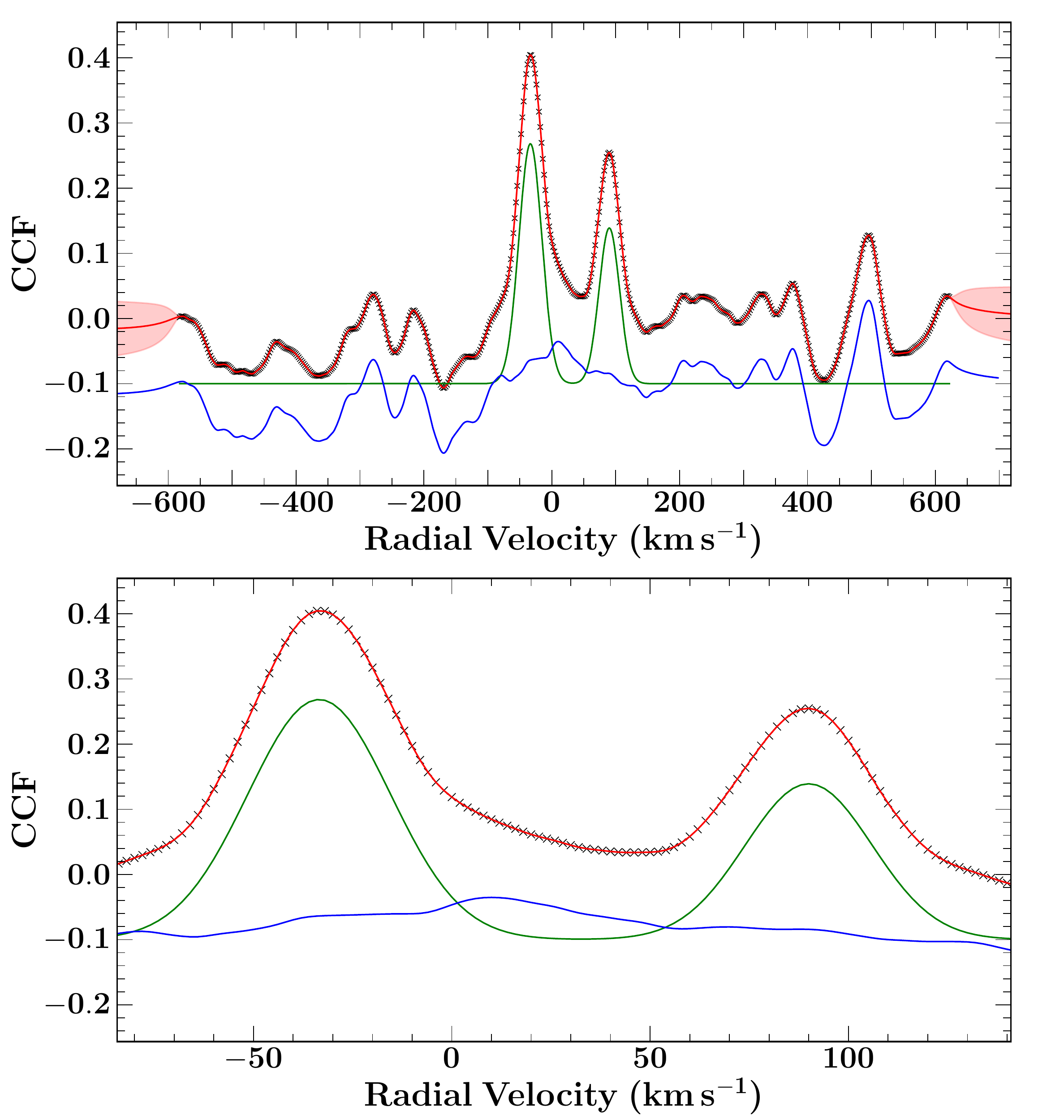} \hfill
\includegraphics[width=0.49\linewidth]{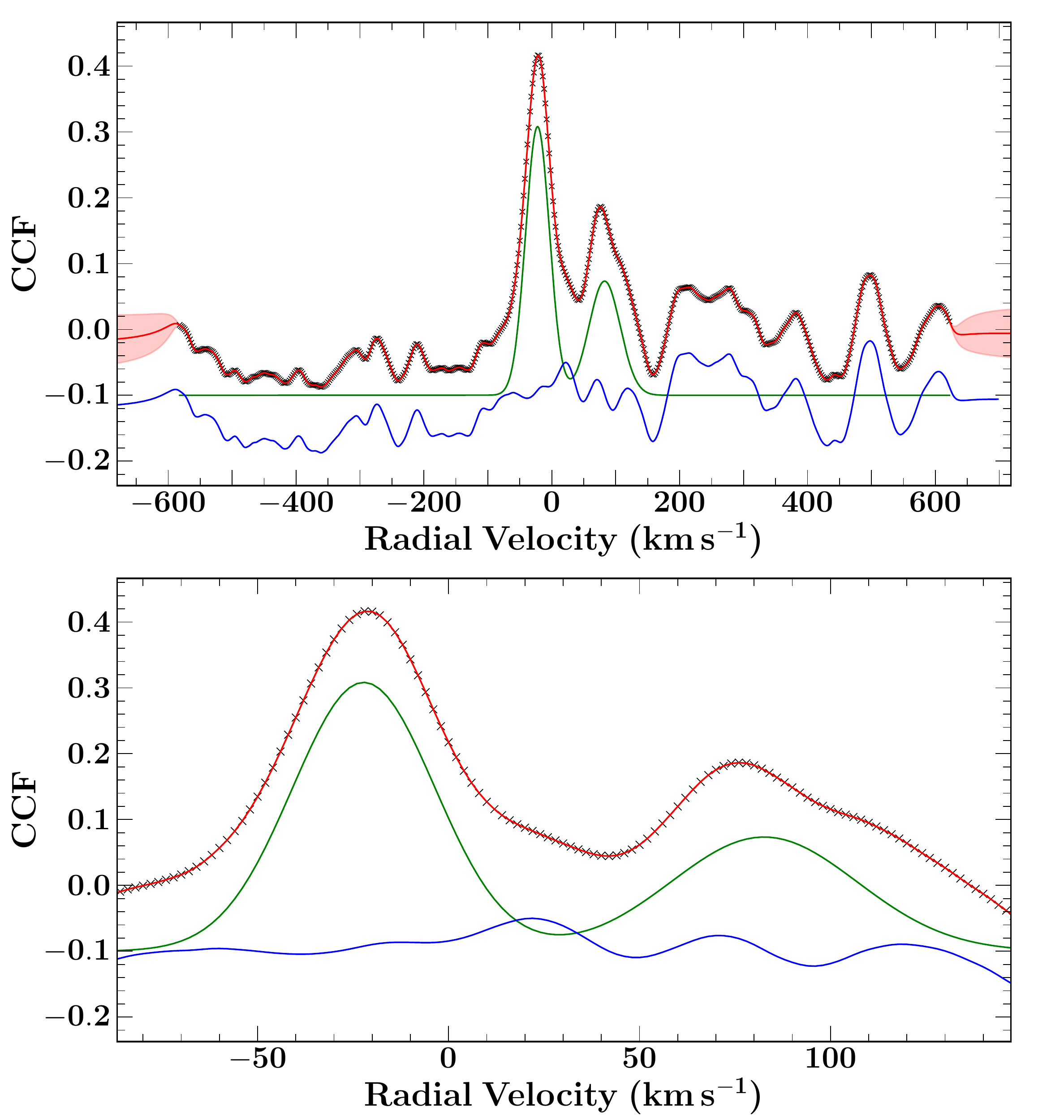}
\caption{Example CCFs (black crosses) obtained from FLAMES spectra taken at HJD = 2455940.6 (left) and 2455979.6 (right), showing the best-fit model (red) together with the associated 68.5\% confidence interval (pink shaded region). The Gaussian and GP terms are also shown separately, in green and blue respectively, with a vertical offset added for clarity. The top panel in each case shows the full range used in the fit and the bottom panel a zoom on the region around the peaks.}
\label{gp_ccf}
\end{figure*}

Before performing the cross-correlation, we masked out regions contaminated by telluric emission. We also masked the calcium infrared triplet in the ISIS spectra, as the lines are very broad and tend to smear out the resulting cross correlation function (CCF). The remaining wavelength intervals used were: 8380--8392, 8455--8484, 8510--8525, 8575--8644 and 8685--8745$\, $\AA\ (see Figure \ref{spec_each}). There were many emission lines present in the FLAMES spectra, not only from the system but also the nebula. These emission lines were not present in the model spectra and inhibited clean cross-correlation. We therefore clipped and interpolated a few data points before and after each emission line to try and fully remove their effect. We also masked the H$\alpha$ emission region (6545--6587\,\AA). The regions used for cross-correlation of the FLAMES spectra were: 6521--6545, 6587--6598, 6610--6642, 6648--6700 and 6702--6713\,\AA. After experimenting with a range of parameters for the MARCS model spectra, we opted to use $T=4000$\,K and $\log g=4$, which gave the best CCF contrast. We did not broaden the template, as that results in broader CCF features which makes it more difficult to separate the peaks corresponding to the two stars. The CCFs were computed and fitted using purpose written {\sc Python} code. 

\begin{table*}
  \centering
  \caption{Radial velocities derived from the WHT/ISIS and VLT/FLAMES spectra. RV values in brackets were not used in deriving the orbital solution, for reasons detailed in the text. $C$ and $W$ are the height and full-width at half-maximum of each peak, respectively.}
  \begin{tabular}{l c c @{\hskip 8mm} c l l @{\hskip 8mm} c l l }
    \hline
    \hline
    \noalign{\smallskip}
    {\bf HJD} & {\bf Orbital}   & {\bf S/N}  &  \multicolumn{3}{c}{{\bf Primary}}  &  \multicolumn{3}{c}{{\bf Secondary}} \\
    & {\bf phase} & (/pixel) & $C$ & $W$  (km\,s$^{-1}$) & RV (km\,s$^{-1}$)  & $C$  & $W$ (km\,s$^{-1}$) & RV (km\,s$^{-1}$)  \\
    \noalign{\smallskip}
    \hline
    \noalign{\smallskip}
    \multicolumn{8}{c}{ISIS} \\  [1.5ex] 
    2455899.51482 & 0.716 & $\sim14$ & 0.36 & $33.3\pm1.8$ & $+73.6\pm1.9$ &
                                      0.28 & $35.4\pm2.5$ & $-68.7\pm2.7$ \\  [1.5ex] 
    2455899.73483 & 0.773 & $\sim13$ & 0.25 & $36.9^{+2.5}_{-3.9}$ & $(+83.4\pm3.9)$ & 
                                      0.08 & $94^{+537}_{-78}$ & $(-57^{+73}_{-236})$ \\  [1.5ex] 
    2455900.55487 * & 0.985 & $\sim19$ & & & &
                                      0.69 & $30.9\pm1.0$ & $(+16.5\pm1.0)$ \\  [1.5ex] 
    2455901.50491 & 0.230 & $\sim18$ & 0.37 & $32.7\pm1.2$ & $(-40.6\pm1.4)$ & 
                                      0.28 & $31.2\pm1.6$ & $(+96.2\pm1.8)$\\  [1.5ex] 
    2455901.57491 & 0.248 & $\sim25$ & 0.60 & $32.7\pm1.1$ & $-50.0\pm1.1$ &
                                      0.42 & $35.0\pm2.0$ & $+95.0\pm1.8$ \\ [1.5ex] 
    2455901.69492 & 0.279 & $\sim24$ & 0.56  & $32.8\pm1.3$ & $-49.2\pm1.2$ &  
                                      0.42 & $33.3\pm1.7$ & $+93.2\pm1.6$ \\ [1.5ex] 
    2455901.77492 & 0.300 & $\sim21$ & 0.51 & $34.8\pm1.3$ & $-49.0\pm1.2$ &
                                      0.43 & $34.8\pm1.6$  & $+88.7\pm1.4$ \\ 
    \noalign{\smallskip}
    \noalign{\smallskip}
    \hline
    \noalign{\smallskip}
    \multicolumn{8}{c}{FLAMES} \\  [1.5ex] 
    2455915.65584 & 0.882 & $\sim26$ & 0.36 & $20.8\pm1.8$ & $+60.5\pm1.6$ &
                                      0.21 & $16.1\pm1.7$ & $-38.1\pm1.7$ \\  [1.5ex] 
    2455917.70423 & 0.411 & $\sim22$ & 0.37 & $19.9\pm1.3$ & $-13.2\pm1.3$ &
                                      0.17 & $34.2^{+6.4}_{-5.4}$ & $+66.5\pm5.3$ \\  [1.5ex] 
    2455918.68222 & 0.663 & $\sim22$ & 0.38 & $17.6\pm1.1$ & $+74.9\pm1.2$ &  
                                      0.15 & $16.1^{+2.1}_{-2.8}$ & $-46.4\pm2.7$ \\  [1.5ex] 
    2455922.69870 & 0.700 & $\sim22$ & 0.40 & $18.2\pm1.1$ & $+75.3\pm1.2$  &
                                      0.20 & $17.9^{+2.1}_{-2.5}$ & $-53.1\pm2.4$ \\  [1.5ex] 
    2455939.65778 & 0.077 & $\sim22$ & 0.37 & $19.0\pm1.6$ & $-8.1\pm1.2$  &
                                      0.18 & $16.7\pm1.6$ & $(+52.6\pm2.2)$ \\  [1.5ex] 
    2455940.64342 & 0.331 & $\sim34$ & 0.38 & $17.6\pm1.1$ & $-33.2\pm1.1$ & 
                                      0.24 & $16.1\pm1.4$  & $+89.9\pm1.5$ \\  [1.5ex] 
    2455940.67976 & 0.341 & $\sim34$ & 0.41 & $18.4\pm1.1$ & $-31.1\pm1.1$ & 
                                      0.27 & $16.2\pm1.2$ & $+88.2\pm1.4$ \\  [1.5ex] 
    2455941.66163 & 0.594 & $\sim35$ & 0.38 & $18.4\pm1.2$ & $+50.9\pm1.0$ & 
                                      0.22 & $24.2\pm3.0$ & $(-19.4\pm2.6)$ \\  [1.5ex] 
    2455943.68750 & 0.117 & $\sim23$ & 0.38 & $18.5\pm1.3$ & $-21.6\pm1.2$ &  
                                      0.19 & $21.4\pm4.0$  & $+76.3\pm3.0$ \\  [1.5ex] 
    2455944.63515 & 0.362 & $\sim32$ & 0.36 & $19.4\pm1.8$ & $-24.9\pm1.5$ &
                                      0.22 & $17.1\pm1.9$ & $+84.4\pm2.1$ \\  [1.5ex] 
    2455945.66720 & 0.628 & $\sim36$ & 0.34 & $16.7\pm1.2$ & $+62.0\pm1.2$ &
                                      0.18 & $16.8\pm1.9$ & $-39.8\pm2.2$ \\  [1.5ex] 
    2455946.68154 & 0.890 & $\sim22$ & 0.36 & $20.7\pm1.6$ & $+56.9\pm1.4$ &  
                                      0.23 & $19.3\pm2.3$ & $-29.1\pm1.9$ \\  [1.5ex] 
    2455977.63971 & 0.880 & $\sim30$ & 0.33 & $19.4\pm1.8$ & $+62.1\pm1.4$ &
                                      0.19 & $20.4^{+3.8}_{-6.0}$ &  $-34.7^{+3.1}_{-4.2}$ \\  [1.5ex] 
    2455979.57946 & 0.380  & $\sim31$ & 0.40 & $18.7\pm1.0$ & $-21.9\pm1.0$ &
                                      0.17 & $26.0^{+6.0}_{-4.8}$ & $(+81.4\pm4.8)$ \\  [1.5ex] 
    2455981.61509 & 0.906  & $\sim28$ & 0.40 & $18.1\pm1.0$ & $+50.5\pm1.0$ & 
                                      0.24 & $24.2\pm2.6$ & $-21.1\pm2.5$ \\                                    
    \noalign{\smallskip}
    \hline
    \label{spec_table}
  \end{tabular}
  \begin{list}{}{}
  \item[*] Only one stellar peak was resolved in the CCF because the spectra was taken close to primary eclipse.
  \end{list}
\end{table*}

We initially performed a simple least-squares fit to the largest peak(s) in the CCFs, using a model consisting of either one or two Gaussians plus a constant offset. However, we found that the resulting RVs were significantly affected by the pronounced correlated noise in the CCFs (examples of which are shown in Figure~\ref{gp_ccf}). This is a major issue when measuring RVs via cross-correlation, particularly for late-type stars. This correlated noise arises in part from noise in the object spectrum, but also -- more critically -- from mismatch between the object and template spectra. A number of approaches have been developed to account for it when evaluating the uncertainties. For example, \citet{Tonry79} decompose the CCF into components which are symmetric and anti-symmetric with respect to the peak, and use the root mean square (RMS) scatter of the antisymmetric component to estimate the CCF noise, on the basis that any `real' CCF signal should be symmetric about the peak. This is the most widely used approach, and it is implemented in the {\sc iraf} cross-correlation package {\sc fxcor}. Working with VLT/FLAMES spectra of moderately faint stars from the OGLE survey, \citet{Bouchy05} used the relation $\sigma_{\rm RV} = 3 \sqrt{W} / S C$, where $\sigma_{\rm RV}$ is the RV uncertainty, $C$ and $W$ are the height and full-width at half-maximum of the CCF peak, respectively, and $S$ is the signal-to-noise of the object spectrum. The form of this relation was deduced from photon-noise considerations, and it was calibrated on multiple observations of a large sample of stars, most of which were presumed to be non-variable. In practice, both of these methods give good results for the objects they were initially designed for, but they tend to systematically under-predict the uncertainties for late-type, rapidly rotating stars.

We therefore decided to model the CCF noise at the same time as fitting for the CCF peaks. Each CCF was modelled as the sum of two Gaussians plus a stochastic noise term, which is described by a GP. The choice of GP kernel was based on a careful examination of the CCFs such as those shown in Figure~\ref{gp_ccf}. Away from the main peak(s), the CCFs display variations on both moderate velocity scales (few tens of m/s) and long velocity scales (few 100 m/s). These cannot be adequately described by a single squared exponential covariance function, which only allows for variations on a single characteristic length scale:
\begin{equation}   
	k_{\rm SE} (r) = A^{2} \exp \left( - \frac{r^{2}}{2l^{2}} \right).
\end{equation}
Instead, we use the rational quadratic:
\begin{equation}   
k_{\rm RQ} (r) = A^{2} \left( 1 + \frac{r^{2}}{2 \alpha l^{2}} \right)^{-\alpha}.
\end{equation}
This can be seen as a squared exponential with a certain amount of additional covariance on large scales, controlled by the parameter $\alpha$ (when $\alpha$ is $>>1$, it reduces to the squared exponential), or alternatively as a scale mixture of squared exponential covariance functions with different characteristic length-scales \citep{Rasmussen06}. We also incorporate a very small white noise term on the diagonal of the covariance matrix to aid convergence. For the sake of computational efficiency, we model the CCFs only in the range $-580 \le v \le 620\,{\rm km}\,{\rm s}^{-1}$, and use only every other data point in the fit. We used a Metropolis-Hastings Markov Chain Monte Carlo (MCMC) with Gaussian proposal distributions to marginalise over the parameters of the Gaussian terms and of the covariance function. We performed five relatively short-chains of 15000 steps, which were sufficient to achieve convergence in all cases; the first 5000 steps were discarded to minimise sensitivity to the initial guesses, and the parameter distributions were derived from the remaining 10000 steps. Figure \ref{gp_ccf} shows example CCFs with the best-fit model obtained in this way.

Not all the CCFs contain well-defined peaks that clearly stand out above the noise. To avoid inferring erroneous RV measurements, we visually inspected every CCF and fit, and discarded any epochs where the contrast of stellar peaks was below $\sim$0.1, or where the properties of the GP component of the fit appeared to change markedly in the region containing the stellar peaks. In a few cases, the best fit to the secondary peak is low compared to the noise, has an unusually large width, and has an RV slightly offset from the expected value (given a preliminary orbital solution based on the rest of the observations). An example of such a case is shown in the right panel of Figure~\ref{gp_ccf}. In such cases, we discarded the secondary measurement, but retained the primary RV. We note that it might be possible to obtain RV measurements from a larger fraction of our spectra by fixing the widths of the two stellar peaks, but we opted not to do this, as we already have enough useful RVs for a good orbital solution, and this would only add data points with large error bars.


\subsubsection{Orbital solution}

\begin{figure}
   \centering
   \includegraphics[width=\linewidth]{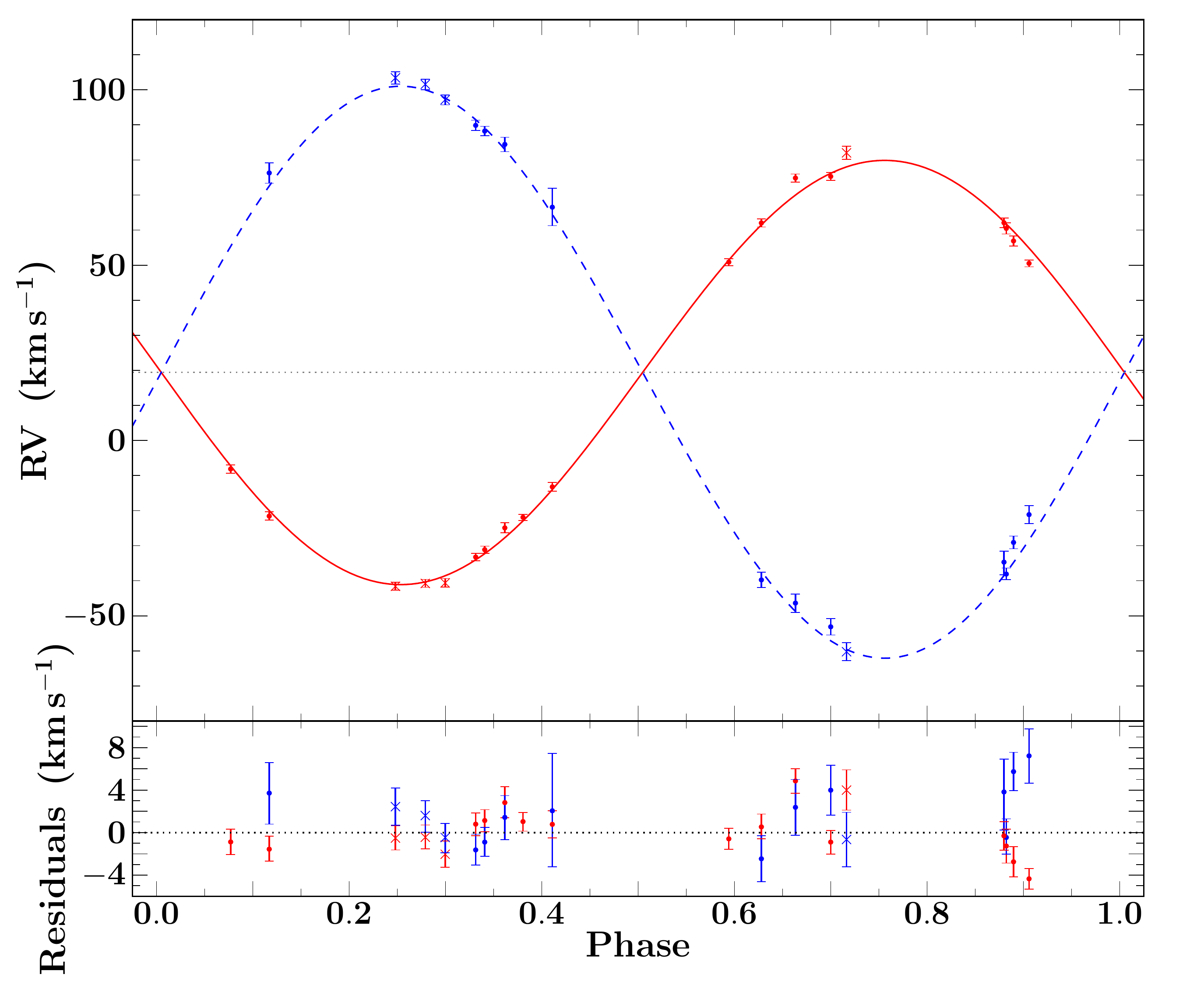}
   \caption{Top: Phase-folded RV data (points with error bars) and best-fit orbital solution for the primary (red solid line) and secondary (blue dashed line). Crosses and circles represent RVs derived from ISIS and FLAMES spectra, respectively.  The grey horizontal dotted line shows the systemic velocity. Bottom: residuals of the best-fit model.}
    \label{rv_curve}
\end{figure}
 
We fit the RV measurements using Keplerian orbits, with free parameters $K_{\rm{pri}}$ and $K_{\rm{sec}}$ (the semi-amplitudes of the primary and secondary), $V_{\rm{sys}}$ (the systemic velocity), as well as the period, $P$, the time of primary eclipse centre, $T_{\rm prim}$, $e \cos \omega$ and $e \sin \omega$. The values and uncertainties obtained from the light curve modelling for the last four of these parameters were used as priors in the RV fit. We used a Metropolis-Hastings MCMC with Gaussian proposal distributions to estimate the uncertainties on the parameters of the RV fit and performed 5 chains of $10^5$ steps each to ensure convergence. Figure~\ref{rv_curve} shows the resulting orbital fit, whose parameters are listed in Table~\ref{params}. The distributions from the MCMC are shown in Figure~\ref{conf_plot} and display negligible degeneracy. 

We find that our systemic velocity of $V_{\rm{sys}} = 19.42 \pm 0.26$\, km\,s$^{-1}$ is consistent with the literature value of the cluster's recessional velocity, $V = 22 \pm 3.5$\, km\,s$^{-1}$ \citep{Furesz06}, suggesting that the system is kinematically associated with the cluster. 
       
\begin{figure}
   \centering
   \includegraphics[width=0.9\linewidth]{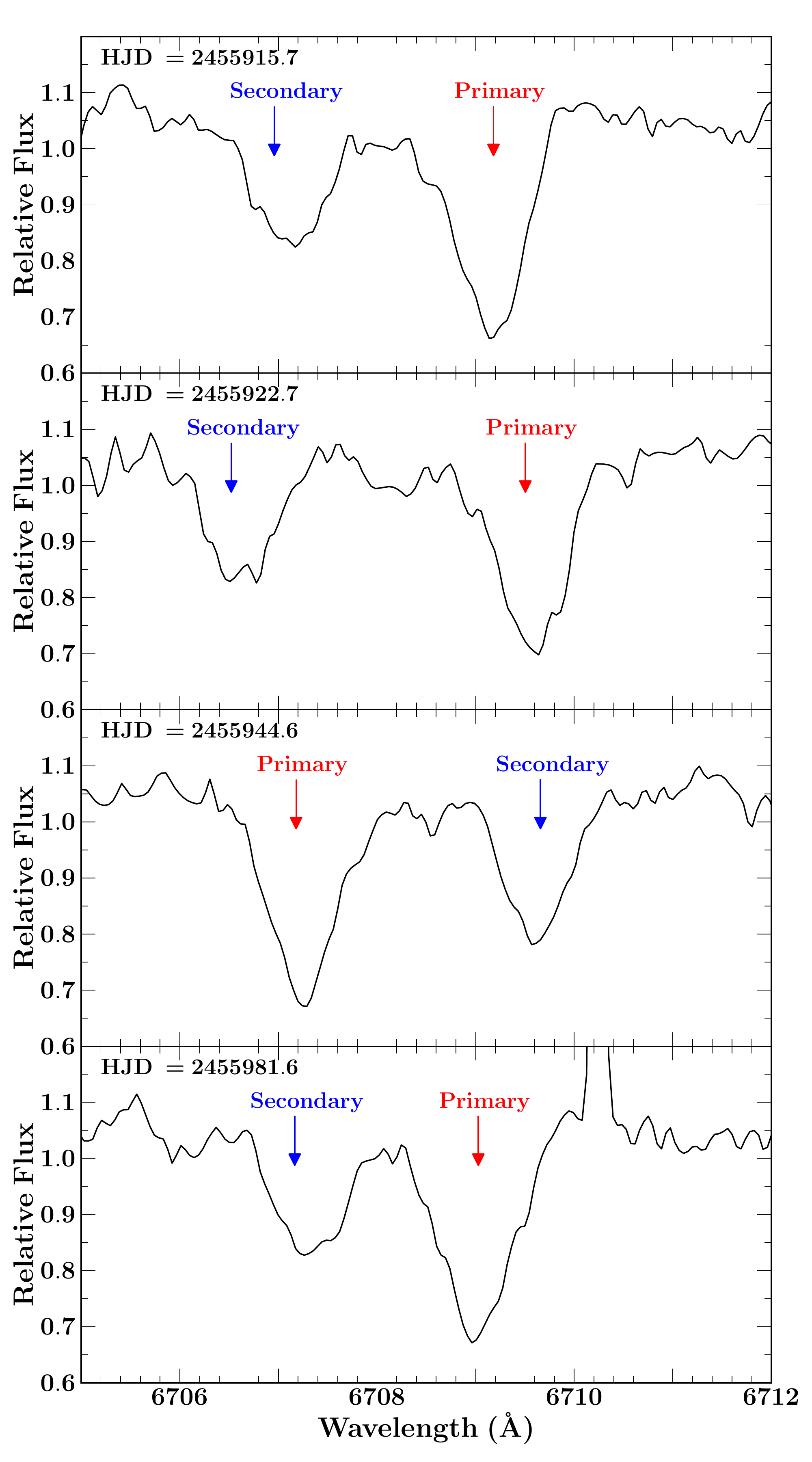}
   \caption{Examples of the lithium 6707.8$\,$\AA\ absorption feature for both primary and secondary stars from FLAMES spectra. From top to bottom, HJD = 2455915.7, 2455922.7, 2455944.6 and 2455981.6. In each case the predicted positions of the two stars are indicated.}
              \label{lithium}
\end{figure}


\subsection{Lithium absorption}

The presence of lithium ($6707.8\,$\AA) absorption in both stars (Figure \ref{lithium}) indicates youth and therefore membership. We determine equivalent widths of both lines by first fitting a cubic spline through the continuum data (masking out all absorption and emission features) and interpolating across each lithium line. We then determine relative flux contributions from each star (using the CCF peak heights for that spectrum) and scale both the data and the spline, before calculating the equivalent widths. We select three spectra in which the lines are well separated and calculate approximate equivalent widths of 0.56 and 0.55\,\AA\  for the primary and secondary, respectively (each individual measurement is mutually consistent with these values to within 0.01 and 0.02\,\AA, respectively). The equivalent widths are very similar, tentatively suggesting that there has not been significant lithium depletion, as the rate of depletion is expected to be faster in the less massive secondary. They are also consistent with other cluster members \citep[e.g.][]{Sergison13}, including early/mid-K spectral types, which are expected to still have their initial abundances at this age.


\subsection{Fundamental Parameters}

From our photometric and radial velocity analysis, we can calculate masses and radii. We propagate our distributions for the period, semi-amplitudes of the velocities, radius sum, radius ratio and inclination to derive masses and radii. These are presented in the bottom part of Table~\ref{params}, along with the surface gravities and orbital separations. The masses of $0.668^{+0.012}_{-0.011}$ and $0.4953^{+0.0073}_{-0.0072}\,M_{\odot}$ are consistent with the combined spectral type derived from the CAHA spectrum. The large radii ($1.295^{+0.040}_{-0.037}$ and $1.107^{+0.044}_{-0.050}\,R_{\odot}$) are what we would expect for a PMS system.

\begin{table*}[p]
  \caption{Fitted and derived parameters of the system.}
  \centering
  \begin{tabular}{l l l l} 
    \hline	
    \hline	
    \noalign{\smallskip}
    Parameter & Symbol & Unit & Value \\
    \noalign{\smallskip}
    \hline
    \noalign{\smallskip}
    \multicolumn{4}{c}{{\sc jktebop} light curve fit} \\
    \noalign{\smallskip}
    \hline
    \noalign{\smallskip}
    Central surface brightness ratio & $J$ & & $0.871^{+0.037}_{-0.035}$ \\  [1.5ex]
    Sum of radii & $(R_{\rm{pri}} + R_{\rm{sec}})/ a$ & &  $0.2198^{+0.0017}_{-0.0018}$ \\  [1.5ex]
    Radius ratio & $R_{\rm{sec}} / R_{\rm{pri}}$ & & $0.854^{+0.058}_{-0.061}$ \\  [1.5ex]
    Fractional primary radius & $R_{\rm{pri}} / a$ & & $0.1186^{+0.0036}_{-0.0033}$ \\  [1.5ex]
    Fractional secondary radius & $R_{\rm{sec}} / a$ & & $0.1013 ~^{+ 0.0039}_{- 0.0045}$ \\  [1.5ex]
    Orbital inclination & $i$ & (\degree) & $85.09  ~^{+  0.16}_{-  0.11}$ \\  [1.5ex]
    Orbital period & $P$ & (days) & $3.8745746\pm0.0000014$ \\  [1.5ex]
    Time of primary eclipse centre & $T_{\rm{prim}}$ & (HJD) & $2454536.76357\pm0.00043$  \\   [1.5ex]
    & $e \cos \omega$  & & $0.00050^{+ 0.00029}_{-0.00028}$ \\  [1.5ex]
    & $e \sin \omega$  & & $-0.0049^{+ 0.0077}_{-0.0075}$ \\
    \noalign{\smallskip}
    \hline
    \noalign{\smallskip}
    \multicolumn{4}{c}{Photometrically constrained RV fit} \\
    \noalign{\smallskip}
    \hline
    \noalign{\smallskip}
    Primary semi-amplitude & $K_{\rm{pri}}$ & (${\rm km}\,{\rm s}^{-1}$) & $60.49 \pm 0.39$   \\    [1.5ex]     
    Secondary semi-amplitude & $K_{\rm{sec}}$ & (${\rm km}\,{\rm s}^{-1}$) & $81.56 \pm 0.62$  \\    [1.5ex]
    Systemic velocity & $V_{\rm{sys}}$ & (${\rm km}\,{\rm s}^{-1}$) & $19.42 \pm 0.26$   \\     [1.5ex]  
    Orbital period & $P$ & (days) & $3.8745745 \pm 0.0000014$ \\   [1.5ex]
    Time of primary eclipse centre & $T_{\rm{prim}}$ & (HJD) & $2454536.76355^{+0.00042}_{-0.00043}$   \\   [1.5ex] 
    & $e \cos \omega$ & & $0.00049^{+ 0.00028}_{- 0.00027}$ \\  [1.5ex]
    & $e \sin \omega$ & & $-0.0033 \pm 0.0040$ \\
    \noalign{\smallskip}
    \hline
    \noalign{\smallskip}
    \multicolumn{4}{c}{Derived parameters}\\
    \noalign{\smallskip}
    \hline
    \noalign{\smallskip}
    Luminosity ratio & $L_{\rm{sec}}/L_{\rm{pri}}$ & & $0.642^{+0.092}_{-0.091}$ \\ [1.5ex]
    Orbital eccentricity & $e$ & & $0.0037^{+ 0.0036}_{-0.0025}$  \\ [1.5ex]
    Semi-major axis & $a$ & ( $R_{\odot}$) & $10.921 \pm 0.056$ \\  [1.5ex]
    Primary mass & $M_{\rm pri}$ & ($M_\odot$) & $0.668^{+0.012}_{-0.011}$ \\ [1.5ex]
    Secondary mass & $M_{\rm sec}$ & ($M_\odot$) & $0.4953^{+0.0073}_{-0.0072}$ \\ [1.5ex]
    Primary radius & $R_{\rm pri}$ & ($R_\odot$) & $1.295^{+0.040}_{-0.037}$ \\ [1.5ex]
    Secondary radius & $R_{\rm sec}$ & ($R_\odot$) & $1.107^{+0.044}_{-0.050}$ \\ [1.5ex]
    Primary surface gravity & $(\log g)_{\rm pri}$ & (cm\,s$^{-2}$) & $4.038^{+ 0.025}_{- 0.026}$ \\ [1.5ex]
    Secondary surface gravity & ($\log g)_{\rm sec}$ & (cm\,s$^{-2}$) & $4.045^{+ 0.040}_{- 0.033}$ \\ [1.5ex]
    Primary semi-major axis$^{\rm a}$ & $a_{\rm{pri}}$ & ($R_\odot$) & $4.651 \pm 0.030$ \\ [1.5ex]
    Secondary semi-major axis$^{\rm a}$ & $a_{\rm sec}$ & ($R_\odot$) & $6.270^{+0.047}_{-0.048}$ \\ 
    \noalign{\smallskip}
    \hline
    \label{params}
  \end{tabular}
  \begin{list}{}{}
  \item[$^{\mathrm{a}}$]relative to centre of mass
  \end{list}
\end{table*}


\section{Discussion}

\subsection{Comparison to pre-main sequence stellar evolution models}

\begin{figure*}
   \centering
   \includegraphics[width=0.9\linewidth]{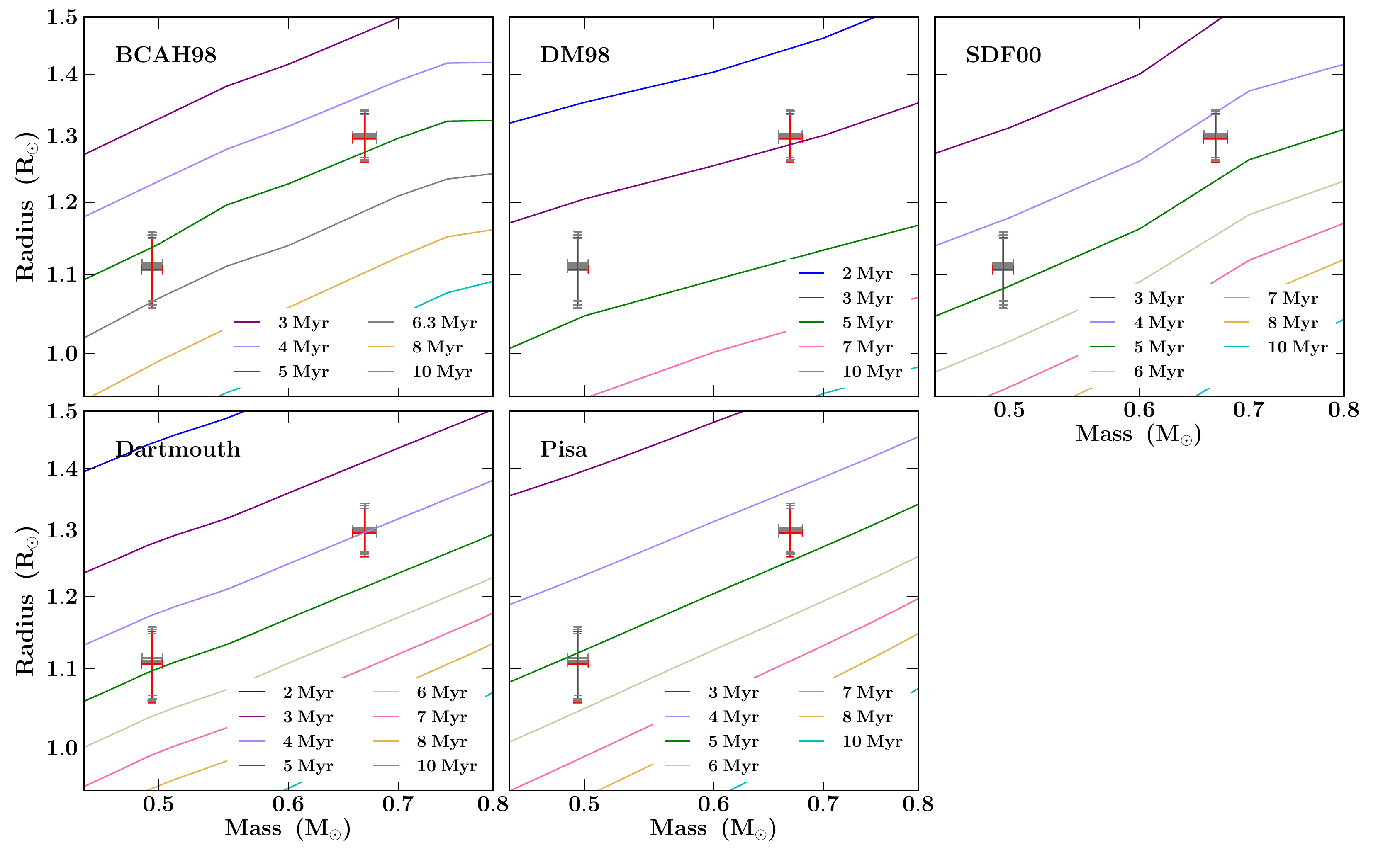}
   \caption{Comparison of measured masses and radii for the two components of CoRoT\,223992193 to the predictions of five sets of theoretical PMS models (see text for details). The different values obtained when varying the limb-darkening coefficients (see Section~\protect\ref{LD_effect}) are shown in grey and the final, adopted values in red.}
   \label{MR_compare}
\end{figure*}

We are now in a position to place the components of CoRoT\,223992193 on the mass-radius diagram (Figure~\ref{MR_plot}, red points). We note that it lies in a very sparsely populated region of the diagram, highlighting its value in testing and constraining PMS stellar evolution models. In Figure~\ref{MR_compare}, we compare the masses and radii of both stars to five sets of widely used pre-main sequence (PMS) isochrones: BCAH98 \citep{Baraffe98}, DM98 \citep{DAntona98}, SDF00 \citep{Siess00}, Dartmouth \citep{Dotter08} and Pisa \citep{Tognelli11}. The parameters of the models shown are: helium mass fraction $Y = 0.282$, metallicity $[ {\rm M}/{\rm H} ] = 0$ relative to solar, and mixing length parameter $\alpha_{\rm{ML}} = 1.9$ for the BCAH98 models; $Y=0.28$, metal mass fraction $Z = 0.02$ and deuterium abundance $X_{\rm D} = 2 \times 10^{-5}$ for the DM98 models; $Y=0.277$ and $Z=0.02$ for the SDF00 models; $Y=0.274$, $Z=0.0189$, $[ {\rm Fe}/{\rm H} ] = [ \alpha /{\rm Fe} ] = 0$ and $\alpha_{\rm{ML}} = 1.938$ for the Dartmouth models and $Y=0.268$, $Z = 0.01$, $X_{\rm D} = 2 \times 10^{-5}$ and $\alpha_{\rm{ML}} = 1.2$ for the Pisa models. These were chosen to match the metallicity of NGC\,2264. For BCAH98 we selected models with a solar-calibrated mixing length $\alpha_{\rm{ML}} = 1.9$, which have been extended down to $0.1\,M_{\odot}$ (I. Baraffe, priv. comm\footnote{The extended models are available at http://perso.ens-lyon.fr/isabelle.baraffe/}). For the Pisa models we chose $\alpha_{\rm{ML}} = 1.2$, as models with lower convective efficiency are known to be more consistent with existing observations of low-mass PMS objects \citep[e.g.][]{Stassun04,Mathieu07}. 

BCAH98, SDF00, Dartmouth and Pisa all manage to fit both components of CoRoT\,223992193 simultaneously. All models under-predict the radius of the primary slightly, compared to that of the secondary. This is a feature which is also seen in other close binaries, and \citet{Chabrier07} proposed that it could be due to enhanced magnetic activity on the primary. However, in the present case the discrepancy is not statistically significant ($<1.5\,\sigma$). More precise masses and radii for this system are needed to distinguish between the different models.

The best match to the BCAH98, SDF00, Dartmouth and Pisa isochrones is obtained for ages $\sim$3.5--6\,Myr. Ages from all isochrones are consistent with the literature estimates of the age of NGC\,2264, given the wide dispersion of the latter. For example, \citet{Dahm08} report a median age of 3\,Myr, but infer an apparent age spread of 5\,Myr from the broadened sequence of cluster members, and \citet{Naylor09} determine a slightly older age of 5.5\,Myr. Various other studies have reported similar ages and large dispersions, all of which depend strongly on the choice of models used to fit the low mass stellar population \citep{Park00}. \citet{Rebull02} found systematic differences in model-derived ages of up to half an order in magnitude for a spectroscopically classified sample of variable stars. Sequential star formation has also been suggested \citep{Adams83}, with the peak rate of low-mass ($M < 0.5\,M_{\odot}$) star formation preceding that of the higher mass population. On the other hand, \citet{Baraffe09,Baraffe12} showed that the broadened cluster sequences observed in a number of star forming regions, which are usually interpreted as arising from an age spread, could also be caused by episodic accretion early on in the star formation process.  Importantly, they predict that, if most of the accreted energy is radiated away (which is expected for the modest initial core masses that will eventually become low-mass stars), accreting stars should have \emph{smaller} radii than non-accreting stars of the same age and mass. Differential accretion history could therefore contribute not only to the apparent age spread within star forming regions such as NGC\,2264, but also to the discrepancies between the parameters of individual components of binary systems such as CoRoT\,223992193. 

In addition, it is worth noting that a non-steady accretion history is also expected to cause enhanced lithium depletion, due to higher temperatures at the base of the convective envelope \citep{Baraffe10}. This is not seen in the present case.


\subsection{Stellar temperatures and the effect of limb darkening}
\label{LD_effect}

When modelling the eclipses, we fixed the limb darkening (LD) coefficients to theoretical values from \citet{Sing10}, given an estimate of the effective temperature of each star. We initially chose temperatures of 4000 and 3750\,K, which maximised the peak of the CCF for the primary and secondary respectively. However, our best-fit masses and radii correspond to somewhat lower effective temperatures, of 3670 and 3570\,K respectively, according to the BCAH98 isochrones. To check whether this has a significant effect on our analysis, we repeated the eclipse modelling using every combination of temperatures for which \citet{Sing10} tabulated LD coefficients, such that the $4250\,{\rm K} \geq T_{\rm eff,pri} \geq T_{\rm eff,sec}$ (LD coefficients were available for 3500, 3750, 4000 and 4250\,K). The results are shown in Figure~\ref{MR_compare}: the values obtained using different combinations of temperatures are shown by the grey symbols, while the adopted values are shown in red. These correspond to effective temperatures of 3750\,K and 3500\,K for the primary and secondary, respectively, and provide the best internal consistency between the effective temperatures used to fix the LD coefficients and those derived from the measured masses and radii according to the BCAH98 isochrones. We note, however, that the masses and radii themselves are essentially insensitive to the choice of LD coefficients, within the range we examined.


\subsection{Distance to CoRoT\,223992193}

Having measured radii and obtained approximate effective temperatures for the two stars, we can evaluate their luminosities and hence the distance to the system. This can then be compared to published estimates of the distance to NGC\,2264, as a further consistency check. We performed this check using both the $V$ and $K$ band magnitudes of the system. The $V$-band magnitude was obtained by averaging the $ugriz$ magnitudes' approximation (according to the prescription of \citealt{Jester05}), and the quoted V-band magnitude from \citet[][V = 16.81]{Dahm07}. This gives $V = 16.74\pm0.1$, where the uncertainty has been inflated to account for the system's variability.

\begin{table} 
  \centering
  \caption{Estimates of the distance to CoRoT\,223992193 using $V$ and $K$-band magnitudes, bolometric corrections \protect\citep[BCs;][]{Bessell98} for different effective temperature estimates, and a range of extinction estimates (from both the literature and as indicated from the CAHA spectrum, Figure~\ref{1039_spec}).}
  \begin{tabular}{llll}
    \hline	
    \hline	
    \noalign{\smallskip}
    Extinction & \multicolumn{2}{c}{ $T_{\rm eff}$ (K)} & Distance \\
    (mag) & (pri) & (sec) & (pc)  \\
    \noalign{\smallskip}
    \hline
    \noalign{\smallskip}
    $A_{V} = 0.71$  &  3500 & 3500  &  $561\pm61$  \\  
    \citep{Dahm05}  &    3750 & 3500  &  $627\pm71$  \\ 
                                  &  3750 & 3750  &  $668\pm73$  \\  [1.5ex]
    
    $A_{V} = 0.45 $ &  3500 & 3500  &  $633\pm69$  \\  
    \citep{Rebull02} &  3750 & 3500  &  $707\pm80$  \\ 
                                  &  3750 & 3750  &  $754\pm83$  \\  [1.5ex]
                                  
    $A_{V} = 0.25$  &  3500 & 3500  &  $694\pm76$  \\
    \citep{Sung97}   &  3750 & 3500  &  $775\pm88$  \\  
                                 &  3750 & 3750  &  $826\pm91$  \\  [1.5ex]	  
    
    $A_{V} = 0.0 $  &  3500 & 3500  &  $776\pm85$  \\  
    (CAHA spectrum) &  3750 & 3500  &  $868\pm98$  \\ 
                                &  3750 & 3750  &  $924\pm101$  \\  [1.5ex]
                                  
    $A_{K} = 0.073$  &  3500 & 3500  &  $766\pm85$  \\
   \citep{Rebull02}  &  3750 & 3500  &  $743\pm82$  \\ 
                                  &  3750 & 3750  &  $728\pm81$  \\  [1.5ex]
    
    $A_{K} = 0.0 $  &  3500 & 3500  &  $791\pm87$  \\  
    (CAHA spectrum) &  3750 & 3500  &  $768\pm84$  \\ 
                                &  3750 & 3750  &  $751\pm83$  \\ 

    \noalign{\smallskip}
    \hline
    \label{dist_tab}
  \end{tabular}
\end{table}

To derive distances to the system, we used the theoretical bolometric corrections (BCs) of \citet{Bessell98}, as they provide the best sampling in our region of the $T_{\rm{eff}}$--$\log g$ parameter space. Empirical BCs are also available in the literature, but these have been shown to give comparable results to the theoretical BCs \citep[e.g.][]{Southworth05}. We adopt the values tabulated for solar metallicity, the closest match available to the metallicity of the cluster, and for $\log g=4.0$, which again provides the best match to our estimate of the surface gravities of both stars. The BCs are tabulated for $T_{\rm eff} = 3500$ and 3750\,K, and are very sensitive to temperature, particularly in the $V$-band, so we computed a combined bolometric magnitude for the system using either value for each star (always ensuring $T_{\rm eff,pri} \geq T_{\rm eff,sec}$). 
Finally, we carried out the calculation for different amounts of extinction along the line of sight, both to the cluster, corresponding to the different determinations available in the literature \citep{Sung97,Rebull02,Dahm05}, and for CoRoT\,223992193, from the CAHA spectrum (Figure~\ref{1039_spec}). The results are reported in Table~\ref{dist_tab}. Note that although we have matched $T_{\rm{eff}}$ and $\log g$ to our estimates for the system, giant stars with the same values would posses different photospheres and so a systematic uncertainty will be present, the effect of which is not well understood.

The V-band extinction estimates in the literature vary widely, mainly because of the different samples of stars used to measure them: the lowest value of $A_V=0.25$  was obtained by \citet{Sung97} using O and B main sequence stars, and the highest ($A_V=0.71$) by \citet{Dahm05} using lower mass H$\alpha$ emitters which may well have dustier environments. \citet{Rebull02} measured an intermediate value of $A_V = 0.45$, as well as $A_K=0.073$, using a spectroscopically selected sample of K and M stars. As the $K$-band distance estimates are less sensitive to reddening, and the BCs in that band less temperature-dependent, we consider the distance estimates we obtain using the $K$-band magnitude of the system and effective temperatures of 3750\,K for the primary and 3500\,K for the secondary, to be the most reliable. We consider the full distance range given by the \citet{Rebull02} determination of $A_{K}$ and that inferred from the CAHA spectrum.

This places the system at a distance of $756\pm96$\,pc, in good agreement with the distances reported for the cluster by \citet{Sung97} ($760 \pm 90$\,pc from isochrone fitting to main sequence B stars) and \citet{Sung10} ($815 \pm 95$\,pc, from fitting the spectral energy distribution of individual members using the models of \citealt{Robitaille07}). \citet{Baxter09} obtained a somewhat larger estimate of $913 \pm 40 \pm 110$\,pc (sampling and systematic errors respectively) by comparing projected rotational velocities and rotational periods, assuming an isotropic inclination distribution. This estimate relies heavily on $v \sin i$ measurements which are notoriously difficult using low signal-to-noise spectra, but is still within $2 \sigma$ of our value for CoRoT\,223992193. Thus we can confirm that the system's magnitude and fundamental parameters are consistent with cluster membership.


\subsection{Spot Modelling}
\label{spot_section}

\begin{figure}
   \centering
   \includegraphics[width=\linewidth]{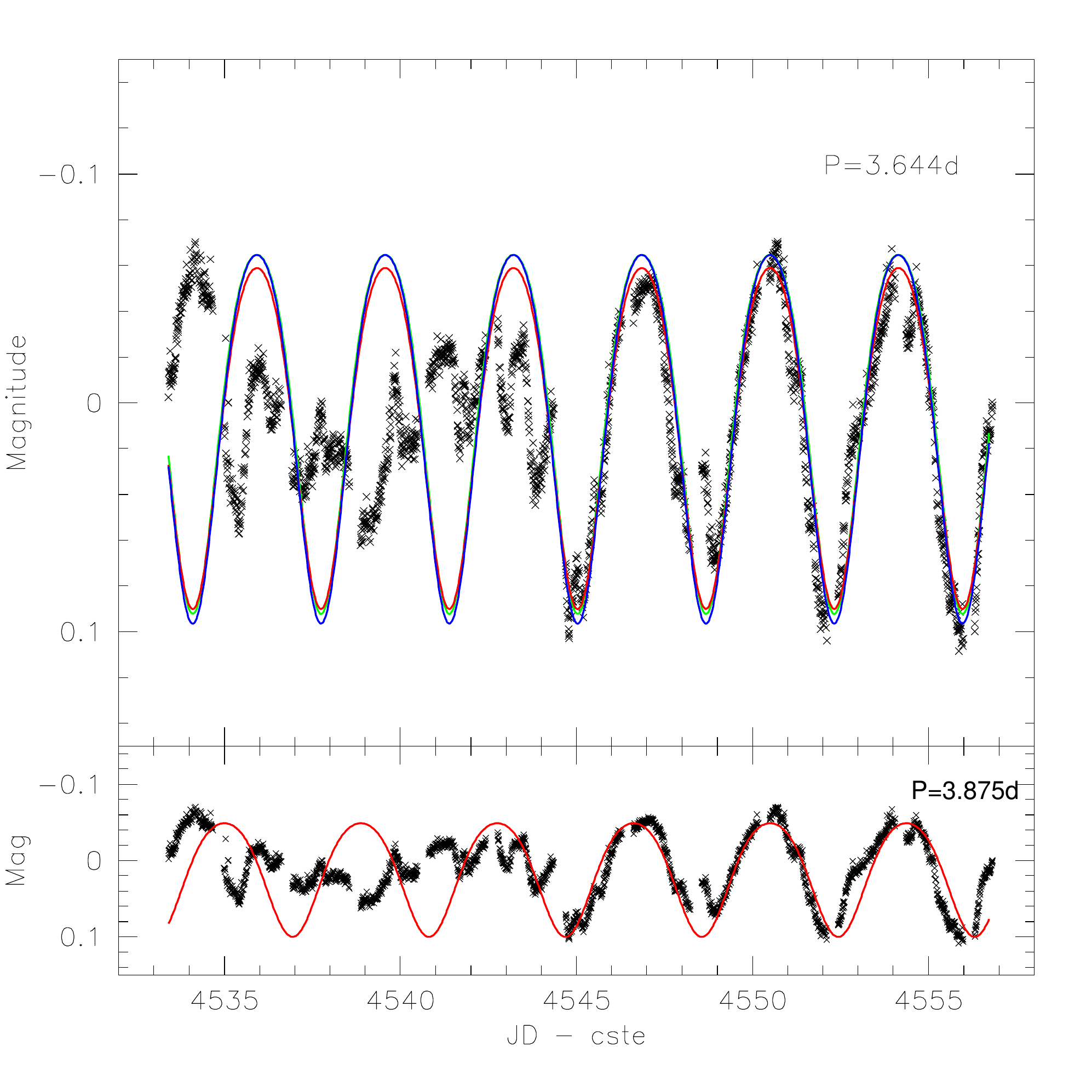}
   \caption{Top: CoRoT 2008 light curve (black crosses) with eclipses removed, showing three versions of our toy spot model (fit to the second half of the light curve only). Red: $T_{\rm{spot}}=3000$\,K, angular radius $=55\degree$ (fraction of stellar surface covered $= 0.21$) and latitude $=70\degree$. Green: $T_{\rm{spot}}=2500$\,K, angular radius $=50\degree$ (fraction of stellar surface covered $= 0.18$) and latitude $=70\degree$. Blue: $T_{\rm{spot}}=3250$\,K, angular radius $=60\degree$ (fraction of stellar surface covered $= 0.25$) and latitude $=65\degree$. Bottom: The red spot model with the period set to the binary orbital period (3.875 days).}
    \label{spot_plot}
\end{figure}

In this section, we investigate the possibility that the OOE variations in  the 2008 CoRoT light curve could be caused by starspots. Young low-mass stars tend to be heavily spotted, and the OOE variations seen in the second half of the light curve, although unusually large in amplitude, have a sinusoidal shape that is not atypical of spotted stars. At the age of NGC\,2264, low-mass stars in close binary systems are expected to be synchronised (due to tidal effects) up to orbital periods of $\sim$7\,days \citep[e.g.][]{Mazeh08}, and indeed the OOE variations in the second half of the light curve appear to be in phase with the eclipses. A Lomb-Scargle periodogram \citep{Horne86} analysis of the second half of the OOE light curve yields a best period of 3.644\,days. The associated formal 1$\sigma$ error is 0.4 days, measured as being the 1$\sigma$ width of a Gaussian fit to the periodogram peak. The OOE period thus appears to be consistent, to within 1$\sigma$, with the orbital period of 3.875 days. However, fitting spot models to the OOE light curve, as shown in Figure~\ref{spot_plot}, reveals that the former period yields much better results than the latter. While the phase of successive minima and maxima are well reproduced using a period of 3.644 days, there is a clear phase drift between the model and the observations when using $P = 3.875$ days.

As a preliminary test of whether starspots could explain the OOE variability, we set up a toy model consisting of a single, large spot located on the primary, facing the secondary, such that flux drops are observed around primary eclipse, as seen in the light curve. The secondary is assumed to be unspotted, and we simply subtract its contribution (39\%) from the total flux. We adopt linear limb-darkening coefficients from \citet{Sing10}, assuming the same temperature for the primary's photosphere ($T_{\rm{pri}} = 3750$\,K) as in the eclipse modelling. We fit three models, with different spot temperatures (2500 to 3250\,K), sizes (60 to 50\,$^\circ$) and latitudes (70 to 60\,$\degree$), to the second half of the light curve. As can be seen in Figure \ref{spot_plot}, each model is able to reproduce the large scale structure of the second half of the light curve. The spots are very large, typically $\sim$20\% of the primary surface, but this is not unusual for young active stars; this active region should more realistically be interpreted as an assembly of smaller spots covering a large fraction of the stellar surface near the pole.

Although the bulk of the large-scale modulations seen in the second half of the light curve can be reproduced by large, cold spots, rapid spot evolution would be required between the first and second halves of the light curve: the active region would essentially have to appear in less than a day. This seems implausible, given its large size \citep[e.g.][]{Grankin08}. In addition, the small scale, short-timescale variability seen throughout the light curve cannot be accounted for by starspots, unless again they evolve very rapidly. Comparing to weak-lined T\,Tauri stars in the 2008 CoRoT observation that show sinusoidal variability, i.e. which should only have spot variability, it is rare to see spot evolution on timescales less than a few weeks. In a few cases we do see evolution, but it is in the form of slow amplitude changes, not erratic day timescale changes. While we cannot rule out spots as a source of variability, it is clear that they cannot be solely responsible for all the variability observed.


\subsection{Spectral energy distribution}
\label{SED_section}

\begin{figure*}
\centering
\includegraphics[width=0.49\linewidth]{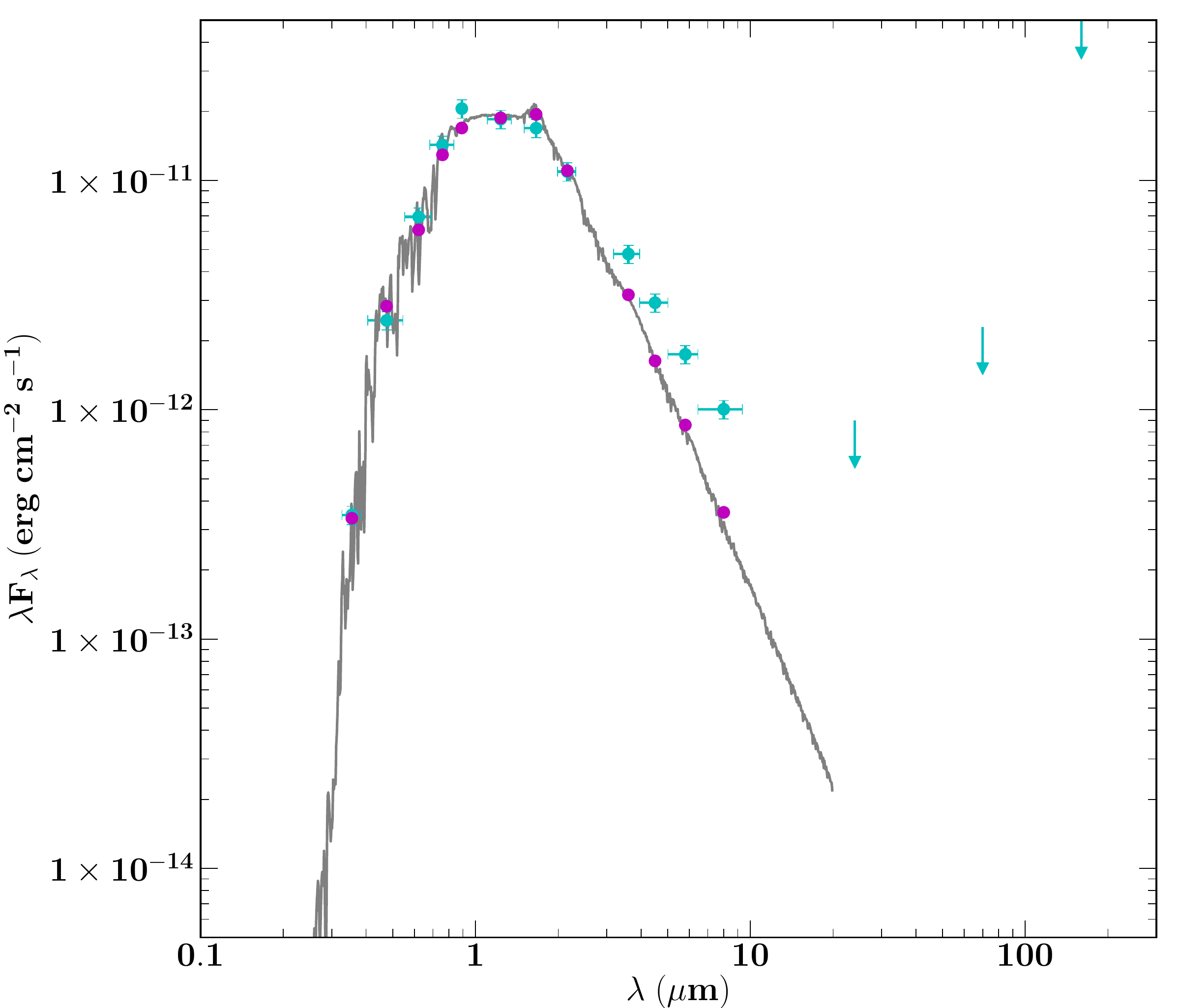} \hfill
\includegraphics[width=0.49\linewidth]{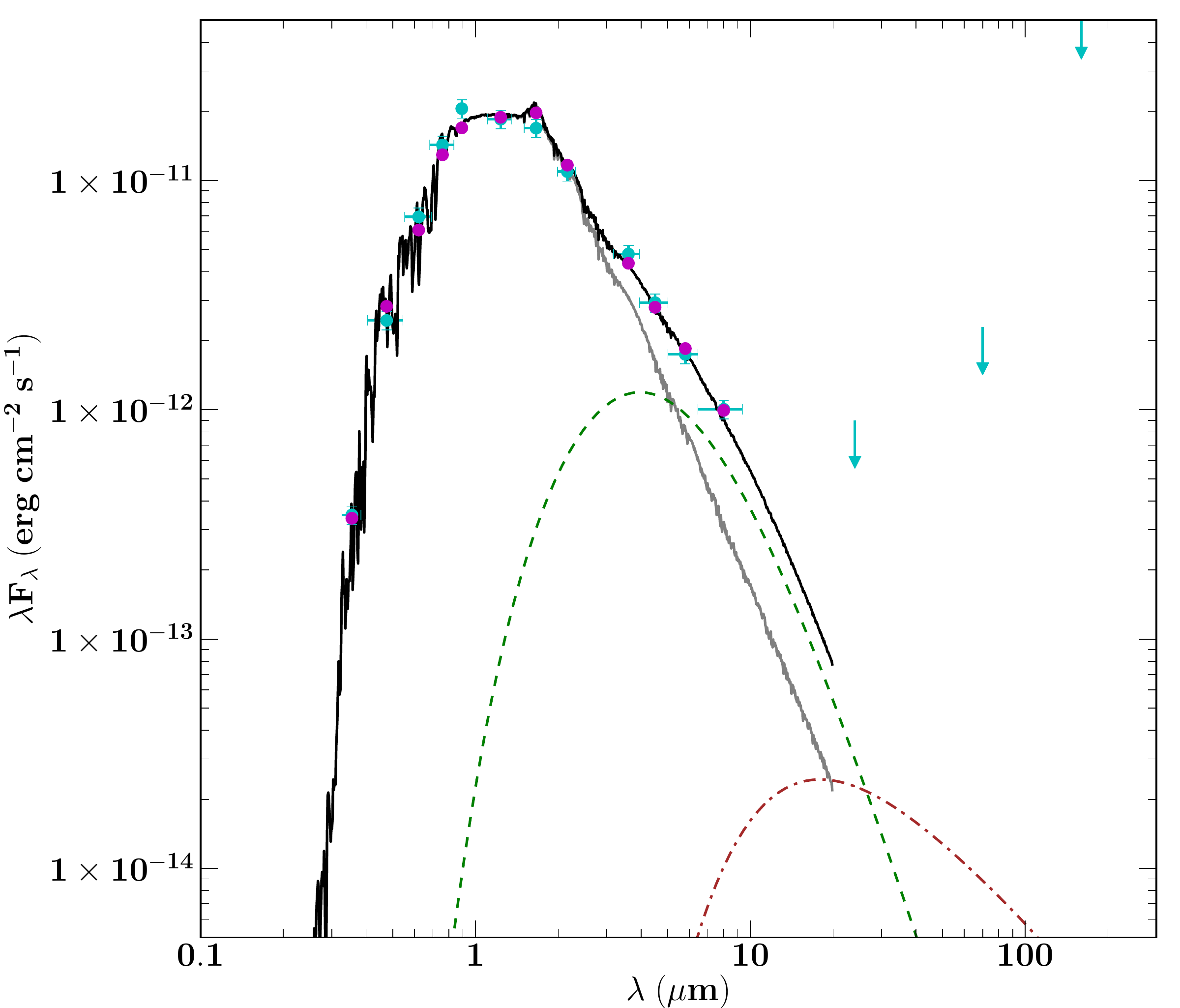}
\caption{Spectral energy distribution of CoRoT\,223992193 (cyan points) based on the magnitudes listed in Table~\protect\ref{eb_props}, plus upper limits in the far-infrared (cyan arrows). Left: the grey line and magenta points show the best-fit two naked photospheres model. Right: the black line and magenta points show the best-fit two naked photospheres model with a small amount of hot dust in the inner cavity of the circumbinary disk (see text for details). The stellar and hot dust terms are shown separately as the grey solid and dashed green lines, respectively. Also shown for completeness, but not used in the fit, is the expected emission from a razor-thin circumbinary disk extending down to 22\,$R_{\odot}$ (brown dot-dashed line), which is illuminated by the central star and heated by the gravitational potential energy released from accretion with $\dot{M} = 10^{-11} M_{\odot}$\,yr$^{-1}$.}
\label{sed}
\end{figure*}

As a likely member of NGC\,2264, the system is presumably no more than a few-to-several Myr old. It is therefore natural to ask whether there is any evidence for circumstellar or circumbinary material in the spectral energy distribution (SED) of the system. This material could conceivably contribute to the large-amplitude, rapidly evolving, out-of-eclipse variability in the CoRoT light curve, particularly as we have seen in Section~\ref{spot_section} that it is difficult to explain this variability using starspots alone. Indeed, the OOE variability of CoRoT\,223992193 is similar to that of classical T\,Tauri stars in the same region \citep{Alencar10}. It is also, to a certain extent, reminiscent of the quasi-periodic flux variations seen of AA Tau \citep[e.g.][]{Bouvier99,Bouvier03,Bouvier07}, which are attributed to occultations of the central star by a warped circum-stellar disk. We therefore constructed an SED for CoRoT\,223992193 using the magnitudes listed in Table~\ref{eb_props}. We also sought evidence of emission at longer wavelengths from archival data, namely \emph{Spitzer}/MIPS 24\,$\mu$m\,\footnote{Spitzer Heritage Archive, \\http://sha.ipac.caltech.edu/applications/Spitzer/SHA/} and \emph{Herschel}/PACS 70 and 160\,$\mu$m\,\footnote{Herschel Science Archive, \\http://herschel.esac.esa.int/Science\_Archive.shtml}. The nebula is prominent at these wavelengths (increasing with wavelength) and displays significant structure. Unfortunately, this system lies on the edge of a nebula filament making detection and analysis more difficult. We performed aperture photometry with Aperture Photometry Tool (APT)\footnote{http://www.aperturephotometry.org/} and found no clear evidence of emission associated with this system. At the PACS wavelengths, the nebula is very bright, and while there is emission above the sky level at CoRoT\,223992193's location, its profile is not Gaussian, it is not centred on the system and its structure largely follows the nebula. At MIPS 24\,$\mu$m, the nebula is less prominent but the emission observed follows the structure of the nebula, and while there is a hint of a slightly larger flux at the system's location, it is not statistically significant given the scatter in the sky flux. A more detailed analysis, accounting for the structured nebula emission would be necessary to determine whether this small excess is real. We therefore attribute the emission at the system's location to the nebula in all cases and compute an upper limit for the flux of CoRoT\,223992193 at each wavelength by taking the `nebula' emission (above that of the median sky level) and quoting the 3$\sigma$ upper limit on this value. These upper limits are: 4.7, 35.2 and 1916.3 mJy at 24, 70 and 160\,$\mu$m respectively. The resulting SED is shown in Figure~\ref{sed}.

A preliminary inspection of the SED suggests that there is a moderate amount of excess flux in the mid-IR. We confirmed this by fitting a stellar photosphere model to the Sloan and 2MASS fluxes, and comparing the best fit model to the observed \emph{Spitzer} fluxes. To do this we constructed a grid of two-photosphere models from pairs of MARCS spectra with $2500 \leq T_{\rm eff} < 8000$\,K, $\log g = 4.0$ and solar metallicity, ensuring that $T_{\rm eff,pri} \geq T_{\rm eff,sec}$, and that the radii of the two stars were within $5 \sigma$ of the measured values. We then convolved the model spectra with the bandpass of each filter and optimised the $\chi^2$ of the fit with respect to the temperatures and radii of the two stars, and the distance and amount of extinction to the system. We constrain the temperatures such that $T \geq 3300$\,K to be consistent with the combined spectral type and temperature ratio from {\sc jktebop}. The best fit parameters are: $T_{\rm{pri}} = 3700$\,K, $T_{\rm{sec}} = 3600$\,K, $R_{\rm{pri}} = 1.42\,R_{\odot}$ (+ 3\,$\sigma$), $R_{\rm{sec}} = 1.15\,R_{\odot}$ (+ 1\,$\sigma$), distance = 830 pc and $A_{V} = 0.1$. The results are shown in the left panel of Figure~\ref{sed}, showing that there is clear excess emission in the mid-IR compared to any reasonable combination of naked stellar photospheres. 

\begin{figure}
   \centering
   \includegraphics[width=\linewidth]{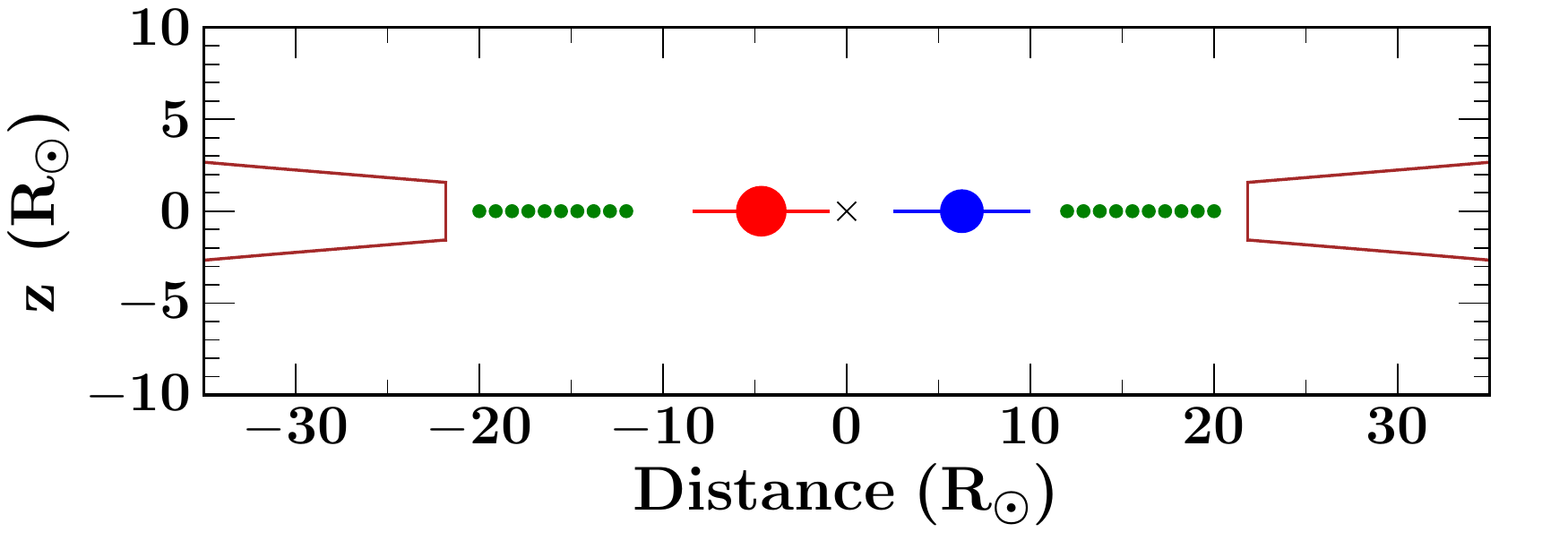}
   \caption{Schematic representation of the proposed system geometry, showing distance from the centre of mass against height above the system plane (z). The primary and secondary stars, along with their circumstellar disks (truncated to a third of the binary separation), are shown in red and blue respectively (the circumstellar disks are shown for completeness but there is no direct evidence for their presence). The sizes and separations of the two stars are to scale. The circumbinary disk (brown) has its inner radius truncated at twice the binary separation. The green dots indicate the general location of dust lying within the inner cavity of the circumbinary disk, such as one could expect to find in accretions streams.}
    \label{sys_view}
\end{figure}

We therefore tested whether this excess could be due to extended dust emission in the vicinity of the two stars. To do this, we must first set out the basic geometry of the system. As the system is young, each star could be  surrounded by a circumstellar disk, and both by a circumbinary disk. Tidal truncation of the circumstellar disks by the other star would limit their outer radii to about one third of the binary separation \citep[][i.e. $\sim$0.017\,AU or 3.6\,R$_\odot$] {Paczynski77,Papaloizou77}. The circumbinary disk would be centred on the centre of mass of the binary, and its inner radius is expected to be roughly twice the binary separation, i.e.\ $\sim$0.1\,AU or $22\,R_\odot$, because the tidal torque exerted by the binary prevents the disk from extending further in \citep{Lin79}. However, material may stream through the inner cavity to be accreted onto the circumstellar disks \citep{Artymowicz96}. A schematic representation of this geometry is shown in Figure~\ref{sys_view}.

There is no evidence for excess emission in the near-IR, which is where we would expect any emission from material in the circumstellar disks to peak. If the primary were a single star, its sublimation radius, i.e.\ the distance beyond which amorphous dust grains typically survive ($T_{\rm{dust~grain}} \leqslant 1500$\,K), would be $3.9\,R_\odot$. This assumes that the temperature of a dust grain at a distance $r$ from a star of luminosity $L$ is given by $T_{\rm dust}=\left[ L/(16 \pi \sigma r^2) \right]^{1/4}$, i.e.\ that the grains are spherical blackbodies, and are located in an optically thin environment. This close to the primary star, the effect of the secondary's flux on the dust temperature is not significant. Similarly, if the secondary star were single, its sublimation radius would be $3.2\,R_\odot$, at which distance the flux of the primary star has a negligible effect. Therefore, the circumprimary disk would not be expected to contain dust and the circumsecondary disk would only be expected to contain a small amount of very hot dust near the sublimation temperature. 
Added to this, the CCF peaks are not obviously rotationally broadened, which given the resolution of the FLAMES spectrograph, 
implies that the two stars have rotational velocities of order 17 km\,s$^{-1}$ or less. Taking this limiting value gives corotation radii (the radii in the 
circumstellar disks where the Keplerian period is equal to the stellar rotation period) of $\gtrsim 8.2$\,$R_{\odot}$ and $\gtrsim 7.4$\,$R_{\odot}$ for the
 primary and secondary stars, respectively. The strong magnetic fields of young, low-mass stars are usually assumed to lead to truncation of the inner disk at 
 or near the corotation radius, which makes it less likely that circumstellar disks exist in this system. Note that the allowed rotational velocities of both stars are consistent with synchronisation; the corresponding rotation periods are $\gtrsim 3.87$ and $\gtrsim 3.31$ days for the primary and secondary respectively.

We therefore attempted to model the mid-IR excess (essentially the residuals of the photosphere-only model shown in the left panel of Figure~\ref{sed} but with the radii of the two stars constrained to within 1\,$\sigma$) as emission from dust in a circumbinary disk. As the inner edge of the circumbinary disk is $>20\,R_\odot$ from the centre of mass of the system, we can approximate the incident flux by that of a single star with luminosity $L_\star=L_1+L_2$, temperature $T_\star$ such that $ \sigma T_\star^4 = L_1/(4 \pi R_1^2) + L_2/(4\pi R_2^2)$ and radius $R_\star$ such that $\sigma T_\star^4 = L_\star/(4 \pi R_\star^2)$. 

We consider a razor thin disk that is aligned with the plane of the binary's orbit and which radiates away energy from both incident stellar flux and the gravitational potential energy released by the accretion of gas. We estimate an order of magnitude accretion rate of $\dot{M} = 10^{-11} M_{\odot}$\,yr$^{-1}$ from the multicomponent H$\alpha$ emission, using the relation described in \citet{Fang09}. This is also consistent with the UV excess upper limit derived from fitting the SED models of \citet{Robitaille07}. We compute the disk temperature as

\begin{equation}
    T_{\rm{disk}} = \left( T_{\rm{irr}}^{\,4} + T_{\rm{acc}}^{\,4} \right)^{1/4} 
\end{equation} 
where, following \citet{Armitage10}, $T_{\rm{irr}}$ and $T_{\rm{acc}}$ are the temperatures arising from stellar irradiation and accretion respectively, and are given by

\begin{equation}
    T_{\rm{irr}}^{\,4} = \frac{T_{\star}^{\,4}}{\pi}  \left(  \sin^{-1} \left( \frac{R_{\star}}{r}  \right) - \frac{R_{\star}}{r} \sqrt{1 - \left( \frac{R_{\star}}{r}  \right)^{2}} ~ \right) 
\end{equation} 

\begin{equation}
    T_{\rm{acc}}^{\,4} = \frac{3 G M_{\star} \dot{M}}{8 \pi \sigma r^{3}} \left( 1 - \sqrt{\frac{R_{\star}}{r}}  ~ \right) 
\end{equation} 

The disk SED is shown by the brown dot-dashed line in the right panel of Figure~\ref{sed}, which is two orders of magnitude smaller than that observed between 3.6 and 8\,$\mu$m.  We need an accretion rate of $\sim$$10^{-7}\,M_\odot$\,yr$^{-1}$ to reproduce the observed fluxes. Such a high accretion rate would lead to much stronger H$\alpha$ emission, and a large UV excess, three orders of magnitude larger than observed (from the $u$-band magnitude of the system). We conclude that the mid-IR excess cannot be explained by emission from a circumbinary disk.

We now consider the possibility that the mid-IR excess is due to optically thin dust located in the inner cavity of the circumbinary disk, following \citet{Jensen97}, who found evidence for this in the SEDs of a number of young spectroscopic binaries. This could arise from ongoing low-level accretion from the circumbinary disk onto the two stars, as predicted by recent numerical simulations \citep{Shi12}. The geometry and kinematics of any dust located within the cavity would, of course, be complex. However, for the purpose of estimating the approximate temperature of the dust and the amount of it needed to reproduce the observed SED, a very simplified toy model is sufficient. We assume that the dust surface mass density varies as $r^{-1/2}$, and that the gas-to-dust ratio is 100.  Again, we approximate the incident radiation field by using a single star with the parameters given above. We are now considering dust located much closer to the two stars, so the validity of this approximation is more doubtful, but it is acceptable for a simple order of magnitude calculation. We model the line-of-sight optical depth through the disk as $\tau_\lambda(r)=\kappa_\lambda \Sigma(r)/ \cos i$, where $\Sigma$ is the gas surface density, $i$ is the angle between the normal to the disk plane and the line of sight, and $\kappa_{\lambda} =  0.1 (\lambda / 250 \, \mu {\rm m} )^{-1}$\,cm$^{2}$\,g$^{-1}$ is the opacity \citep{Jensen97}.  The emitted flux is then given by:
\begin{equation}   
      \lambda F_\lambda =   \frac{\cos i}{D^2} \int_{r_{\rm in}}^{r_{\rm out}} \lambda B_\lambda   [T_{\rm dust}(r)] \left( 1 - e^{-\tau_\lambda(r)} \right) 2 \pi r \,dr.
\end{equation}  
where the inner and outer radii of the cavity, $r_{\rm in}$ and $r_{\rm out}$, should be larger than the sublimation radius of the single star we consider ($5\,R_\odot$) and similar to the circumbinary disk's inner radius, respectively. The flux received from the cavity increases with the mass of dust as long as $\tau_\lambda <1$ and becomes insensitive to this mass when the optical depth reaches unity\footnote{Even if the dust is optically thin when considering light from the star incident on a particle within the cavity, it could be optically thick when viewed (near) edge-on, i.e.\ from the point of view of an observer on Earth.}.  We find that we can explain the observed mid-IR excess with as little as $\sim$$1 \times 10^{-13}\,M_\odot$, which gives $\tau_{\lambda}$\,$\sim$1 at the outer edge of the cavity for $\lambda$ between 3.6 and 8\,$\mu$m. In this model, the cavity extends from $\sim$\,5 to 32\,$R_\odot$ with corresponding dust temperatures of $\sim$\,1450 to 600\,K. The resulting fit to the SED is shown in the right panel of Figure~\ref{sed}. Although the outer radius of the dust model is slightly larger than the expected inner edge of the circumbinary disk, they are of the same order of magnitude, which is sufficient given the very simplified nature of our model. Note that the mass of dust required and the extent of the cavity are insensitive to interstellar reddening (up to $A_V=1.0$). As a consistency check, we compared the mass of dust we calculate to that required to reproduce the estimated mass accretion rate, and find them in agreement to within an order of magnitude.

Our model is very crude, but it does demonstrate that the observed SED can be reproduced by invoking a very small amount of dust within the inner cavity of a circumbinary disk, such as one could expect to find in accretion streams. Unfortunately, the flux measurements beyond 10$\,\mu$m are only rather weak upper limits, because of emission by interstellar dust filaments (nebulosity) superimposed on the target, so we cannot place any constraints on the circumbinary disk itself. However, the presence of such a disk would certainly be required to replenish the dust in the cavity. This is because when dust grains enter the inner cavity, their motion is essentially ballistic; the accretion timescale is therefore close to the free fall time.


\section{Conclusions and future work}

We report a new double-lined, detached eclipsing binary, which is comprised of two pre-main sequence M-dwarfs, discovered by the CoRoT space telescope in the NGC\,2264 star forming region. We have measured the fundamental parameters of both stars using the continuous 23.4 day light curve obtained by CoRoT in March 2008, as well as 22-epochs of radial velocity data obtained almost four years later with VLT/FLAMES and WHT/ISIS. The orbit is consistent with circular ($e = 0.0037^{+ 0.0036}_{-0.0025}$) and has a period of $3.8745745 \pm 0.0000014$\,days and a separation of $10.921 \pm 0.056\,R_{\odot}$.  The primary and secondary stars have masses of $0.668^{+0.012}_{-0.011}$ and $0.4953^{+0.0073}_{-0.0072}\,M_{\sun}$, and radii of $1.295^{+0.040}_{-0.037}$ and $1.107^{+0.044}_{-0.050}\,R_{\sun}$, respectively. The systemic velocity is within $1\sigma$ of the cluster median which, along with the presence of lithium absorption, strongly indicates cluster membership. 

This is only the ninth PMS EB system with component masses below $1.5\,M_\odot$, and it lies in a region of the mass-radius plane where existing observational constraints are very scarce. Within the current uncertainties, the parameters of the two stars are essentially consistent with the predictions of PMS stellar evolution models for ages of $\sim$3.5--6\,Myr. Although we have broken the degeneracy between the radius and surface brightness ratios, which can be a severe limitation in grazing EB systems, the final uncertainties on the component masses and radii are still a few percent. As highlighted by \citet{Torres10}, sub-percent accuracies are needed to place truly significant constraints on evolutionary models. We hope to improve the constraints on this system in the future, by incorporating additional data and refining our treatment of the out-of-eclipse variability.

Modelling of the system's broadband optical and infrared SED reveals a mid-IR excess that cannot be explained by emission from two stellar photospheres alone; additional, cool material is required. Dynamical arguments and simple disk models indicate that neither circumstellar nor circumbinary disks can explain this excess. We find that the excess can be reproduced by dust emission from within the inner cavity of a circumbinary disk and show that only a very small dust mass of $\sim$$1 \times 10^{-13}$\,$M_{\odot}$ is required. Such small amounts of dust could be found in accretion streams from a circumbinary disk.
This opens up the possibility that some of the OOE variations may be due to obscuration of the central stars by dust located at the inner edge, or in the central cavity, of the circumbinary disk.

NGC\,2264 was re-observed by CoRoT in December 2011 and January 2012, as part of a co-ordinated multi-observatory campaign involving \emph{Spitzer}, \emph{Chandra}, MOST\footnote{Note that CoRoT\,223992193 fell outside the \emph{Chandra} and MOST fields of view} (Microvariability and Oscillations of STars) and a host of ground-based observatories, including CFHT $ugr$ photometry and VLT/FLAMES spectroscopy\footnote{The FLAMES spectra used to derive RVs in the present paper were obtained as part of this campaign.}. Although the sampling of the observations by other telescopes is significantly sparser than that of the CoRoT data, the widely separated bandpasses of the CFHT, CoRoT and warm \emph{Spitzer} data should help to disentangle the contributions of starspots and any occultations by dusty material in the out-of-eclipse variability. We will also model the time-dependent H$\alpha$ emission seen in the FLAMES spectra, which may shed light on any ongoing accretion in the system. 

A number of other EBs were discovered in the field of NGC\,2264 during the same CoRoT observation. Once we have characterised those which belong to NGC\,2264 in more detail, they will form a unique sample of near-coeval EB systems formed from the same molecular cloud, and may shed further light on the range of ages and accretion histories in this region.

\begin{acknowledgements}

We thank Isabelle Baraffe for extending her models of stellar evolution (with mixing length $\alpha = 1.9$) down to lower masses allowing us to compare the source's masses and radii to these model predictions.
We thank D. Barrado, the Director of Calar Alto Observatory, for granting Discretionary Time to this project.
We also thank Catarina Alves de Oliveira for providing her automatic fitting method to derive the spectral type of the source from the Calar Alto low resolution spectrum. 

This research is based on data collected by the CoRoT satellite, which is publicly available via the IAS Data Center at {\tt http://idoc-corot.ias.u-psud.fr/}.
Based on observations made with ESO Telescopes at the La Silla Paranal Observatory under programme ID 088.C-0239(A).
The WHT and INT are operated on the island of La Palma by the Isaac Newton Group in the Spanish Observatorio del Roque de los Muchachos of the Instituto de Astrof'sica de Canarias.  
Based on observations collected at the Centro Astron—mico Hispano Alem‡n (CAHA) at Calar Alto, operated jointly by the Max-Planck Institut fŸr Astronomie and the Instituto de Astrof'sica de Andaluc'a (CSIC). 
This work is based in part on observations made with the Spitzer Space Telescope, obtained from the NASA/ IPAC Infrared Science Archive, both of which are operated by the Jet Propulsion Laboratory, California Institute of Technology under a contract with the National Aeronautics and Space Administration.  
This work is based in part on observations made with \textit{Herschel}, which is an ESA space observatory with science instruments provided by European-led Principal Investigator consortia and with important participation from NASA.
This publication makes use of data products from the 2MASS, which is a joint project of the University of Massachusetts and the Infrared Processing and Analysis Center/California Institute of Technology, funded by the National Aeronautics and Space Administration and the NSF. 
This research has made use of the SIMBAD data base, operated at CDS, Strasbourg, France.
Funding for the SDSS and SDSS-II has been provided by the Alfred P. Sloan Foundation, the Participating Institutions, the National Science Foundation, the U.S. Department of Energy, the National Aeronautics and Space Administration, the Japanese Monbukagakusho, the Max Planck Society, and the Higher Education Funding Council for England. The SDSS Web Site is {\tt http://www.sdss.org/}, where the list of the funding organizations and collaborating institutions can be found.

This work was supported in part by the UK Science and Technology Facilities Council through grant ST/G002266 (SA)
and studentship ST/J500641/1 (EG).
AM acknowledges support from a Science and Technology Facilities Council studentship.
JB thanks CNES for the partial funding of this project. 
SHPA acknowledges financial support from CNPq, CAPES and Fapemig. 
JS acknowledges funding from STFC in the form of an Advanced Fellowship.
Finally, we would like to thank the anonymous referee for their careful reading of the manuscript and helpful suggestions for improvement.

\end{acknowledgements}

\bibliographystyle{aa}
\bibliography{ref} 

\end{document}